\documentclass[nonacm,acmsmall,screen]{acmart}

\usepackage[]{hyperref}

\usepackage{booktabs}   
\usepackage{subcaption} 

\usepackage{comment}
\usepackage{makecell}
\usepackage{graphicx}
\usepackage{wrapfig}
\usepackage{xspace}
\usepackage{amsmath}
\usepackage{amsfonts}
\usepackage{amsthm}
\usepackage{bm}
\usepackage{textcomp}
\usepackage{stmaryrd}
\usepackage{mathtools}
\usepackage{enumerate}
\usepackage{enumitem}
\usepackage{multirow}
\usepackage{array}
\usepackage{wrapfig}
\usepackage{xcolor}
\definecolor{clcommon}{RGB}{245,245,245}
\definecolor{clctrl}{RGB}{218,232,252}
\definecolor{clconc}{RGB}{225,213,231}
\definecolor{cldata}{RGB}{213,232,212}
\definecolor{clguard}{RGB}{255,242,204}
\definecolor{clundstd}{RGB}{176,227,230}
\definecolor{clopt}{RGB}{250,215,172}
\definecolor{clbug}{RGB}{248,206,204}
\definecolor{clsec}{RGB}{186,200,211}
\definecolor{clcommon-dark}{RGB}{102,102,102}
\definecolor{clctrl-dark}{RGB}{108,142,191}
\definecolor{clconc-dark}{RGB}{150,115,166}
\definecolor{cldata-dark}{RGB}{130,179,102}
\definecolor{clguard-dark}{RGB}{214,182,86}
\definecolor{clundstd-dark}{RGB}{14,128,136}
\definecolor{clopt-dark}{RGB}{180,101,4}
\definecolor{clbug-dark}{RGB}{184,84,80}
\definecolor{clsec-dark}{RGB}{35,68,93}
\definecolor{vkeyviolet}{HTML}{7030a0}
\usepackage{listings}
\definecolor{vgray}{RGB}{100,100,100}
\definecolor{vgreen}{RGB}{104,180,104}
\definecolor{vblue}{RGB}{75,105,198}
\definecolor{vorange}{RGB}{255,143,102}
\lstdefinestyle{mystyle}{
	commentstyle=\color{vgreen},
	keywordstyle=\bfseries\color[HTML]{8F00FF},
	numberstyle=\tiny\color{vgray},
	stringstyle=\color{vorange},
	basicstyle=\ttfamily\footnotesize,
	breakatwhitespace=false,
	breaklines=true,
	captionpos=b,
	keepspaces=true,
	numbers=left,
	numbersep=5pt,
	showspaces=false,
	showstringspaces=false,
	showtabs=false,
	tabsize=4,
    xleftmargin=4.0ex,
    numbersep=1.5ex
}
\DisableLigatures[>,<]{encoding=*,family=tt*}

\lstdefinestyle{vstyle}{
    basicstyle=\small\ttfamily,
    keywordstyle=\color{vblue},
    columns=fullflexible
}

\setlist[enumerate]{leftmargin=2em}
\setlist[itemize]{leftmargin=1em}

\newcommand{\mylink}[2]{\href{#2}{#1}\footnote{\url{#2}}}

\newcommand{\veriX}{\texttt{X}\xspace}
\newcommand{\aname}[1]{\texttt{#1}}
\newcommand{\acategory}[1]{\emph{#1}}

\usepackage{tikz}

\usetikzlibrary {arrows.meta}
\DeclareRobustCommand\deparrow{\scalebox{0.9}{\raisebox{-2pt}{\tikz{
    \draw[line width = 1pt, arrows = {-Kite[length=5pt,width=4pt,inset=1pt]}] (0,0) -- (0, 0.35);
    \draw[line width = 1pt] (-0.08, 0.07) -- (0.08, 0.07);}}}}
\DeclareRobustCommand\deparroww{\scalebox{0.9}{\raisebox{-2pt}{\tikz{
    \draw[line width = 1pt, arrows = {-Kite[length=5pt,width=4pt,inset=1pt]}] (0,0) -- (0, 0.35);
    \draw[line width = 1pt] (-0.08, 0.07) -- (0.08, 0.07);
    \draw[line width = 1pt] (-0.08, 0.14) -- (0.08, 0.14);}}}}
\DeclareRobustCommand\deparrowww{\scalebox{0.9}{\raisebox{-2pt}{\tikz{
    \draw[line width = 1pt, arrows = {-Kite[length=5pt,width=4pt,inset=1pt]}] (0,0) -- (0, 0.35);
    \draw[line width = 1pt] (-0.1,0) -- (0, 0.18);
    \draw[line width = 1pt] (0.1,0) -- (0, 0.18);}}}}

\newcommand{\msym}[1]{\mathit{#1}}
\newcommand{\mterm}[1]{\textcolor{vblue}{\textbf{\texttt{#1}}}}
\newcommand{\mclass}[1]{\textcolor{vgreen}{\texttt{#1}}}
\newcommand{\mcomment}[1]{\textcolor{gray}{\text{#1}}}
\newcommand{\mverb}[1]{\mathtt{#1}}
\newcommand{\mite}[3]{#1\ \mathtt{?}\ #2\ \mathtt{:}\ #3}
\newcommand{\reset}{\mverb{reset}}
\newcommand{\acc}{\mverb{acc}}
\newcommand{\iin}{\mverb{in}}
\newcommand{\uundefined}{\mverb{undefined}}

\newcommand{\Entry}{\text{Entry}}

\newcommand{\Cycle}{\text{Cycle}}

\usepackage[commandnameprefix=ifneeded,commentmarkup=footnote]{changes}
\definechangesauthor{cql}
\definechangesauthor[color=red]{yue}
\definechangesauthor[color=blue]{tt}
\definechangesauthor[color=red]{TODO}
\definechangesauthor[color=red]{CHK}

\DeclareRobustCommand{\hasse}{
    \ \ \raisebox{-0.42\height}[0.5\height]{
        \footnotesize
        \begin{tikzpicture}[scale=0.8,thick]
            \node (0) at (-0.6, 0)    {$\mathtt{0}$};
            \node (1) at (+0.0, 0)    {$\mathtt{1}$};
            \node (B) at (-0.3, +0.9) {$\mathtt{B}$};
            \node (U) at (+0.3, -0.9) {$\mathtt{U}$};
            \node (X) at (0.6, 0)     {$\mathtt{X}$};
            \node (Z) at (1.2, 0)     {$\mathtt{Z}$};
            \node (T) at (0.3, 1.8)   {$\mathtt{T}$};

            \draw[->,arrows={-Latex[width=0pt 3,length=3pt]},scale=0.5] (U) -- (X);
            \draw[->,arrows={-Latex[width=0pt 3,length=3pt]},scale=0.5] (U) -- (Z);
            \draw[->,arrows={-Latex[width=0pt 3,length=3pt]},scale=0.5] (U) -- (0);
            \draw[->,arrows={-Latex[width=0pt 3,length=3pt]},scale=0.5] (U) -- (1);
            \draw[->,arrows={-Latex[width=0pt 3,length=3pt]},scale=0.5] (0) -- (B);
            \draw[->,arrows={-Latex[width=0pt 3,length=3pt]},scale=0.5] (1) -- (B);
            \draw[->,arrows={-Latex[width=0pt 3,length=3pt]},scale=0.5] (X) -- (T);
            \draw[->,arrows={-Latex[width=0pt 3,length=3pt]},scale=0.5] (Z) -- (T);
            \draw[->,arrows={-Latex[width=0pt 3,length=3pt]},scale=0.5] (B) -- (T);
        \end{tikzpicture}
    }
    \ \
}

\begin{document}

\title{Qihe: A General-Purpose Static Analysis Framework for Verilog}

\author{Qinlin Chen}
\orcid{0009-0006-5498-5927}
\email{qinlinchen@smail.nju.edu.cn}
\affiliation{
  \institution{Nanjing University}
  \country{China}
}

\author{Nairen Zhang}
\orcid{0009-0004-3737-8768}
\email{nairen@smail.nju.edu.cn}
\affiliation{
  \institution{Nanjing University}
  \country{China}
}

\author{Jinpeng Wang}
\orcid{0009-0007-9099-4262}
\email{jpwang@smail.nju.edu.cn}
\affiliation{
  \institution{Nanjing University}
  \country{China}
}

\author{Jiacai Cui}
\orcid{0009-0001-4922-887X}
\email{jiacaicui@smail.nju.edu.cn}
\affiliation{
  \institution{Nanjing University}
  \country{China}
}

\author{Tian Tan}
\orcid{0009-0009-3792-1237}
\authornote{Corresponding author.}
\email{tiantan@nju.edu.cn}
\affiliation{
  \institution{Nanjing University}
  \country{China}
}

\author{Xiaoxing Ma}
\orcid{0000-0001-7970-1384}
\email{xxm@nju.edu.cn}
\affiliation{
  \institution{Nanjing University}
  \country{China}
}

\author{Chang Xu}
\orcid{0000-0002-6299-4704}
\email{changxu@nju.edu.cn}
\affiliation{
  \institution{Nanjing University}
  \country{China}
}

\author{Jian Lu}
\orcid{0009-0002-1166-6927}
\email{lj@nju.edu.cn}
\affiliation{
  \institution{Nanjing University}
  \country{China}
}

\author{Yue Li}
\orcid{0009-0009-1285-2298}
\authornotemark[1]
\email{yueli@nju.edu.cn}
\affiliation{
  \department{and all authors are also affiliated with the State Key Laboratory for Novel Software Technology}
  \institution{Nanjing University}
  \country{China}
}


\begin{abstract}
In the past decades, static analysis has thrived in software, facilitating applications in bug detection, security, and program understanding.
These advanced analyses are largely underpinned by general-purpose static analysis frameworks, which offer essential infrastructure to streamline their development.
The generality of these frameworks is often founded on components such as a front end for program conversion, an intermediate representation (IR), and a suite of fundamental analyses providing key program information.
Soot exemplifies such a framework in Java, sparking thousands of analyses and impacting both academia and industry.
Conversely, hardware lacks such a framework, which may limit the exploration of diverse static analysis applications in this field.
This gap could overshadow the promising opportunities for sophisticated static analysis in hardware, hindering achievements akin to those witnessed in software.
To advance static analysis in hardware, we introduce Qihe, the first general-purpose static analysis framework for Verilog---a highly challenging endeavor given the absence of precedents in hardware.
Qihe features an analysis-oriented front end, a Verilog-specific IR, and a suite of diverse fundamental analyses that capture essential hardware-specific characteristics---such as bit-vector arithmetic, register synchronization, and digital component concurrency---and enable the examination of intricate hardware data and control flows.
These fundamental analyses are specifically designed to support a wide array of analysis clients for hardware.
To validate Qihe's utility, we further developed a series of analysis clients spanning bug detection, security, and program understanding, and applied them to real-world hardware projects.
Unlike software, real-world hardware projects tend to contain fewer but harder-to-detect bugs, as they typically undergo extensive simulation and rigorous verification to prevent the prohibitive costs of hardware defects.
Despite this, our preliminary experimental results are highly promising;
for example, Qihe uncovered 9 previously unknown bugs in popular real-world hardware projects (averaging 1.5K+ GitHub stars), all of which were subsequently confirmed by developers; moreover, Qihe successfully identified 18 bugs beyond the capabilities of existing static analyses for Verilog bug detection (i.e., linters), and detected 16 vulnerabilities in real-world hardware programs.
These results underscore the transformative potential of static analysis in hardware design.
By open-sourcing \mylink{Qihe}{https://qihe.pascal-lab.net}, which comprises over 100K lines of code, together with its carefully curated and comprehensive \mylink{documentation}{https://qihe-docs.pascal-lab.net}, we aim to inspire further innovation and applications of sophisticated static analysis for hardware, aspiring to foster a similarly vibrant ecosystem that software analysis enjoys.
\end{abstract}





\maketitle 

\section{Introduction} \label{sec:intro}

In the software domain, static analysis is widely recognized as an effective technique for enhancing software quality.
Over the years, numerous sophisticated static analyses have been developed to achieve various application tasks, such as detecting bugs~\cite{pldi06chord}, identifying security vulnerabilities~\cite{flowdroid}, and aiding in program understanding~\cite{ecoop16slicing}.

Despite its proven successes in software, static analysis remains underutilized in hardware, particularly with Verilog, the predominant hardware description language (HDL)~\cite{snug2016lang}.
The current situation mirrors the early days of software static analysis, where its applications were limited to compiler-integrated optimizations and basic syntax checks (e.g., early linting tools), with more sophisticated software analyses, such as those for bug detection and security, yet to emerge.
In hardware, static analysis techniques are primarily adopted by hardware synthesizers (e.g., Yosys~\cite{yosys}) for compiler optimization purposes;
linting tools~\cite{slang,verible,svlint,verilator} are popularly used to enforce code-style or syntactic checks but lack the sophistication needed to detect deeper design flaws like improper register resets and hardware Trojans.

We argue that one of the main reasons for the stagnation of hardware static analysis is the lack of \emph{a general-purpose framework} that provides the necessary infrastructure to enable the development of sophisticated analyses.
In the software domain, such frameworks have proven transformative.
For example, Soot~\cite{soot}, a general-purpose static analysis framework for Java, has catalyzed thousands of analysis works in both academia and industry.
Research efforts like FlowDroid~\cite{flowdroid}, widely used for detecting security vulnerabilities in Android apps, rely on Soot's fundamental analyses for identifying control and data flows statically.
Some companies also leverage FlowDroid and Soot to develop their own analyses, which allows them to identify bugs early in the development cycle, leading to potentially significant cost savings.

Drawing from the designs of existing frameworks for software analysis~\cite{soot,wala,taie}, a \emph{general-purpose} framework typically requires the close cooperation of the following key components:
(1)~\emph{fundamental analyses} that provide basic program information used across various application-specific analyses, 
(2)~\emph{a front end} and \emph{an IR} that derive an analysis-oriented program abstraction from source code, and
(3)~\emph{an analysis manager} that facilitates the reuse of results from existing analyses and the integration of new ones.

However, given the absence of a general-purpose framework in hardware, the development of hardware static analysis typically relies on one of the following three kinds of options, each with its own limitations.
Linters such as Slang~\cite{slang}, Verible~\cite{verible}, SVLint~\cite{svlint}, and Verilator-Lint~\cite{verilator} are confined to syntax- or pattern-based analyses on abstract syntax trees (ASTs).
Tools like Pyverilog~\cite{pyverilog} and VeriPy~\cite{veripy} provide hardware abstractions beyond ASTs but lack the infrastructure to address critical hardware features like synchronization and concurrency, highly limiting their utility for diverse analysis needs.
While CIRCT~\cite{circt}, a comprehensive hardware compiler framework primarily used for synthesis, can also be adapted for static analyses, it lacks the fundamental facilities necessary for building sophisticated analyses that require intricate value flows and deep semantic reasoning, resulting in substantial development costs. 

In this paper, we present Qihe, the first general-purpose static analysis framework for Verilog, comprising three key components:

\begin{itemize}

\item \emph{A Set of Fundamental Analyses} (Section~\ref{sec:analyses}):
We have developed 22 fundamental analyses to statically construct the hardware program information necessary to support a wide array of analysis clients.
This unique contribution stems from our decade-long experience in frontline research and development in foundational static software analysis, sustained communication with industrial hardware practitioners, and an in-depth understanding of Verilog semantics.
 
\item \emph{An Analysis-Oriented IR and Front End} (Section~\ref{sec:ir}):
We have designed the IR and front end to work collaboratively, meeting specific analysis needs such as supporting analysis for incomplete hardware programs and retaining additional source information, while still ensuring the IR remains simple for ease of analysis.

\item \emph{An Analysis Manager} (Section~\ref{sec:manage}):
We have designed an engineering-focused mechanism that enables new developers to conveniently integrate and execute new analyses within Qihe.

\end{itemize}

To validate Qihe as a general-purpose framework capable of supporting a variety of application-specific analyses, we developed 20 client analyses, which also serve as examples to guide future developers in creating their own.
These analyses target common but non-trivial needs identified through surveys on hardware bugs and vulnerabilities~\cite{asplos22bugbed, cwe,hackdac18,hackdac21,iccd13trusthub,jhss2017trusthub} along with insights from industrial hardware practitioners,
spanning diverse application domains, including bug detection, security analysis, and program understanding.

We conducted experiments by applying these client analyses to a set of real-world Verilog programs, including System-on-Chip (SoC) designs, encryption modules, communication protocols (e.g., AXI implementations), and more~\cite{asplos22bugbed-artifact,iccd13trusthub,jhss2017trusthub,xiangshan,darkriscv,verilogprojects,basicverilog,zipcpu}.
The preliminary experimental results are summarized as follows:

\begin{itemize}
\item
\emph{Hardware Bug Detection}:
Unlike software, hardware tends to contain fewer but harder-to-detect bugs, since it typically undergoes extensive simulation and rigorous verification to avoid the prohibitive costs of defects~\cite{dvcon16cost, dac04costs}.
Despite this, our analyses uncovered 9 previously unknown bugs in popular real-world hardware projects (averaging over 1.5K GitHub stars), all of which were subsequently confirmed by the developers.
Examples include registers not correctly reset in a RISC-V CPU, an unreachable state in a CRC generator, unexpected bit truncation in a SHA512 encryption module, etc.
Our analyses also successfully detected another 9 hardware bugs: some were collected from commit histories of real-world programs by previous surveys, while others were summarized by industrial hardware experts and injected into real programs as benchmarks~\cite{asplos22bugbed,hackdac18,hackdac21}.
Note that all 18 bugs, spanning eight categories as summarized in Table~\ref{tab:bugs}, cannot be detected by existing static analyses for Verilog bug detection~\cite{slang,svlint,verible,verilator}.
\item
\emph{Hardware Security Analysis}:
Our analyses detected 15 vulnerabilities related to information leakage, particularly involving the exposure of secret encryption keys via observable output ports like antennas and capacitance.
Additionally, a hardware Trojan~\cite{dac17mux} is also detected that exploits the special \veriX value in Verilog for concealment.
\item
\emph{Hardware Program Understanding}:
Our analysis helps developers understand the semantics of Verilog synthesis by predicting their physical implementation details, such as whether a variable is mapped to a clock signal, latch, or register, with high recall and precision.
For example, to minimize latency, clocks should avoid being used in complex logic in the code, but these clock usages are often identified after a time-consuming synthesis process. 
With static analysis, these issues can be identified earlier, such as during the development phase, thereby reducing potential high engineering costs.
\end{itemize}

Moreover, from the preliminary experimental results, we highlight three complementary points. 

Firstly, existing static analysis techniques for hardware struggle to detect the aforementioned bugs and vulnerabilities due to their inability to reason about complex value flows within hardware programs.
This underscores the necessity for sophisticated analyses to uncover deep hardware issues, highlighting the role of Qihe as a foundational framework for advancing hardware analysis.

Secondly, the detection of these deep hardware issues has traditionally relied on resource-intensive methods such as model checking and fuzzing, as well as labor-intensive techniques like simulation-based testing~\cite{pldi11caisson,asplos15secverilog,fmcad07ic3,tcad24modelchecking,usenixsecurity22fuzzing1,usenixsecurity22fuzzing2}, which often dominate and slow down the hardware development process. 
In contrast, static analysis can be deployed at an earlier stage and is often considerably more lightweight. For instance, on a real-world RISC-V SoC codebase comprising 1.8 million lines, Qihe can complete all 20 client analyses, along with 22 fundamental analyses, in approximately 5 minutes using 40GB of memory. 

Finally, the development cost for these client analyses is significantly reduced thanks to the infrastructure provided by the Qihe framework, particularly its extensive suite of fundamental analyses.
On average, each client analysis requires only about 300 lines of Java code when implemented leveraging Qihe's infrastructure, whereas developing the same analysis from scratch---without utilizing Qihe's fundamental analyses---would require over 5,000 lines of code on average.

In summary, this paper makes a singular contribution by introducing Qihe, the first general-purpose static analysis framework for Verilog, consisting of over 100K lines of Java code.
Qihe is open-source and can be obtained from its \mylink{website}{https://qihe.pascal-lab.net}.
We also provide comprehensive \mylink{documentation}{https://qihe-docs.pascal-lab.net} to support the development of diverse application-specific analyses.

The growth of hardware analysis may not mirror the widespread success of software analysis. Yet, over 25 years ago, it was also hard to foresee that static analysis would help enhance software quality so significantly.
Qihe is a bold venture, uncovering the initial potential of sophisticated hardware analysis. We hope it serves as a catalyst for collaboration between software and hardware communities to jointly shape the still-blurry yet promising future of hardware analysis.

\section{Overview of Qihe}

\begin{figure}[t]
\centering
\includegraphics[width=\linewidth]{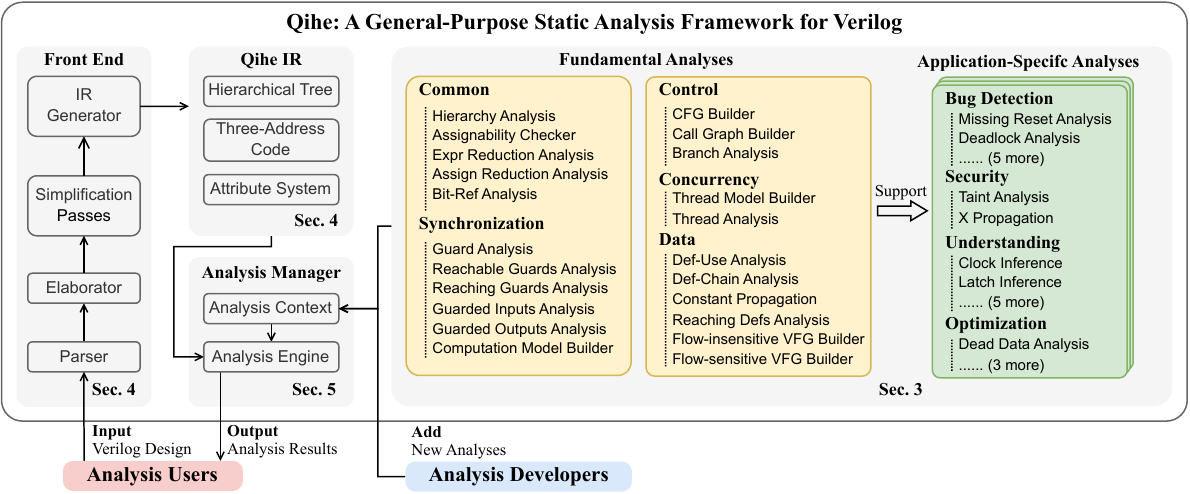}
\caption{Workflow of Qihe. It supports both \emph{analysis users} in executing analyses on Verilog designs and \emph{developers} in creating new analyses, enabled by the collaboration of its components.
}
\label{fig:arch}
\end{figure}

Qihe, as a general-purpose static analysis framework for Verilog, is designed to address two key use cases: enabling \emph{analysis users} to execute available analyses on their hardware designs and empowering \emph{analysis developers} to create diverse new analyses using the Qihe framework.
Figure~\ref{fig:arch} illustrates the underlying workflow for each use case, supported by four core components: the front end, Qihe IR, diverse analyses, and the analysis manager.
Below, we describe this workflow in detail, guided by the figure.
Further details about Qihe's components are discussed in later sections.

For analysis users, the process begins by inputting Verilog designs into the \emph{front end}, which converts them into \emph{Qihe IR}, a program abstraction for Verilog.
The \emph{analysis manager} then executes the request analyses, as configured by users, on the Qihe IR and delivers the results to users.
Specifically, the manager detects available analyses and configurations, registers them into an analysis context, and invokes an analysis engine to apply them to the Qihe IR.
Analyses are executed in topological order based on their dependency relations to ensure correctness. 

For analysis developers, the \emph{analysis manager} provides an intuitive interface for creating new analyses (whether fundamental or application-specific) based on existing ones.
Once new analyses are implemented, the manager automatically detects and registers the new analysis into the analysis context, making it available for users to execute.
We encourage developers to leverage Qihe to explore and contribute more potentially useful analyses.

\section{Analysis Suite}\label{sec:analyses}

\begin{figure}[t]
\centering
\includegraphics[width=\linewidth]{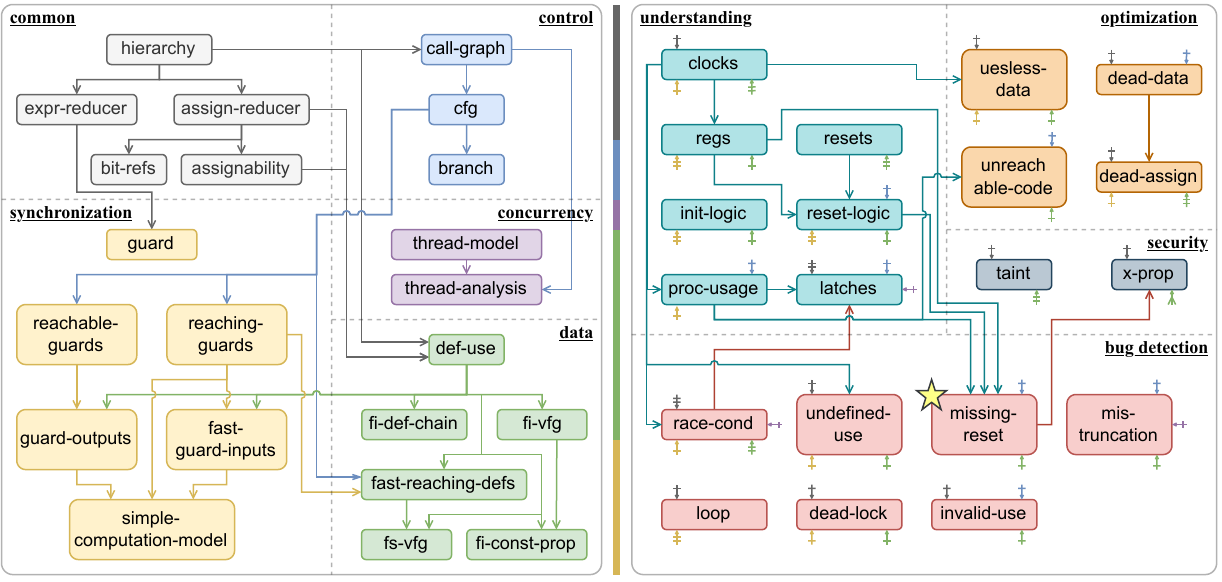}
\caption{
Analyses and their dependencies in Qihe.
Each analysis is represented by its short name. 
The left section presents fundamental analyses, categorized into five groups, while the right section illustrates application-specific analyses across four domains.
Arrows depict dependency relations between these analyses. To maintain clarity, we use short arrows (``\deparrow'', ``\deparroww'', and ``\deparrowww'') to succinctly represent the number of dependencies that an application-specific analysis has on fundamental analyses. 
These arrows denote dependencies on one, two, or three fundamental analyses within a specific category, with the category identified by the arrow's color (or position in black-white printing).
For example, the starred analysis \texttt{missing-reset} has two short arrows ``\deparrow'': one top-right and one bottom-right, signifying its dependency on \emph{one} analysis from the \acategory{control} category and \emph{one} from the \acategory{data} category.
We use the central bar to visualize the relative usage of each category by color and length.}
\label{fig:analyses}
\end{figure}

Qihe supports the development of diverse sophisticated analyses by utilizing the cooperation among various analyses.
Figure~\ref{fig:analyses} depicts an overview of all the major analyses in Qihe and their dependencies, illustrating such cooperation.
Note that the dependencies between fundamental analyses and client analyses are represented by the special arrows described in the figure caption.

Guided by this figure, Section~\ref{sec:analysis-overview} first explains how the analyses are derived and categorized.
Section~\ref{sec:motivating-example} then elaborates on \aname{missing-reset}, a representative bug detection client in Qihe, to demonstrate the sophistication required to detect real-world hardware bugs that elude existing linter-style static analyses for Verilog bug detection~\cite{svlint,slang,verible,verilator} and highlight the fundamental differences between hardware analysis and typical software analysis.
Finally, Section~\ref{sec:supporting-case-study} discusses how existing analyses in our framework collaborate to support this critical hardware bug detection client.\looseness=-1

\subsection{Overview of All Analyses}\label{sec:analysis-overview}

Qihe offers a rich suite of fundamental and application-specific analyses.
Since no general-purpose static analysis framework for Verilog exists as a reference, we derived these analyses through a three-step process that leverages surveys, research experience, and engineering practices:
(1) We identified analysis needs by surveying hardware bugs and security issues~\cite{asplos22bugbed,cwe,hackdac18,hackdac21,iccd13trusthub,jhss2017trusthub}, examining existing analyses in synthesis and linting tools~\cite{circt,yosys,svlint,slang}, and directly communicating with industrial hardware practitioners~\cite{hisilicon,t-head};
(2) Building on this survey and leveraging our decade-long research experience in fundamental static analysis for software, we distilled \emph{22 fundamental analyses} providing essential Verilog program information commonly required across diverse application needs;
and (3) To demonstrate how to build diverse applications by leveraging fundamental analyses in Qihe, we additionally developed \emph{20 application-specific analyses} to address common but non-trivial needs identified in our survey.

To facilitate a clear understanding of these analyses, we categorize them by their design purpose, as shown in Figure~\ref{fig:analyses}.
Fundamental analyses (on the left) are grouped into five categories based on their focus on specific aspects of hardware design: \acategory{common} language features, \acategory{data} and \acategory{control} flows, and \acategory{concurrency} and \acategory{synchronization} characteristics.
Application-specific analyses (on the right) are categorized into four popular application domains: hardware \acategory{bug detection}, \acategory{security} vulnerability identification, program \acategory{understanding}, and program \acategory{optimization}.

To enhance analysis modularity and reusability, each analysis in Qihe is designed with a single purpose and delegates other responsibilities to upstream analysis as dependencies.
This practice naturally forms intricate dependencies, as depicted in Figure~\ref{fig:analyses}.

For a deeper understanding of these analyses and their dependencies, we next elaborate on a representative analysis in Qihe (Section~\ref{sec:motivating-example}) and then discuss its supporting dependencies (Section~\ref{sec:supporting-case-study}).\looseness=-1

\subsection{Sophisticated Analysis for Hardware: A Motivating Example}\label{sec:motivating-example}

This section presents a motivating example that illustrates three key questions:
(1) What do hardware bug-detection problems look like, and how do they differ from their software analogues?
(2) How to design sophisticated hardware static analyses that leverage hardware-specific insights to address these problems?
(3) What is the challenge in building sophisticated hardware static analysis?
\looseness=-1

Our motivating example is the \aname{missing-reset} analysis, which detects a class of critical hardware bugs where registers fail to initialize properly during circuit reset, creating unstable system states that attackers could potentially exploit~\cite{unreset,tifs24llm,asplos22bugbed,tcad22rtlcon}.
It serves as a representative Verilog analysis in our analysis suite that not only handles diverse hardware-specific semantics---such as clocks, registers, and resets---that are uncommon in software analysis, but also demonstrates the level of sophistication required to detect real-world hardware bugs (summarized in Table~\ref{tab:bugs}) that elude existing linter-style static analyses for Verilog bug detection~\cite{svlint,slang,verible,verilator}.

Below, we first introduce the \emph{missing-reset} problem in Verilog (Section~\ref{sec:motivation-problem}) and distinguish it from the classical \emph{use-before-initialization} (UBI) problem in software (Section~\ref{sec:motivation-vs}).
We then present the core insights underlying our \aname{missing-reset} analysis for addressing this problem (Section~\ref{sec:motivation-analysis}).
Finally, we use this analysis to discuss the key challenge in developing sophisticated hardware analyses (Section~\ref{sec:motivation-challenges}).

\subsubsection{The Missing-Reset Problem in Verilog}\label{sec:motivation-problem}

The \emph{missing-reset} problem occurs when hardware \emph{registers} lack proper \emph{reset}.
Below, we first provide the necessary Verilog background on \emph{registers} and \emph{reset}, and then present the \emph{problem formulation}.

\begin{wrapfigure}{r}{5.36cm}
    \includegraphics[trim=0cm 3.78cm 8.52cm 0,clip,page=2]{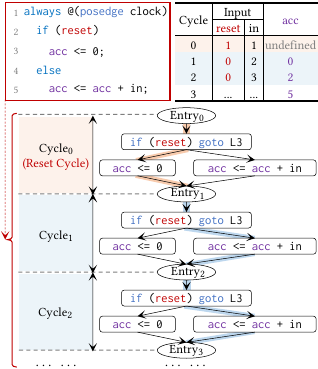}
    \caption{A properly-reset register example.}\label{fig:background}
    \vspace{-1em}
\end{wrapfigure}
\emph{Registers.}
A hardware register is a digital circuit component that stores and updates values under the control of a driving clock signal alternating between 0 and 1;
Verilog uses variables and \texttt{always} blocks to model hardware registers.
As exemplified in the code snippet of Figure~\ref{fig:background}, $\acc$ is a hardware register whose driving clock is specified in the \texttt{always} block header (line~1) and whose value update logic is specified in the block body (lines~2--5).
Functionally, this Verilog program describes an accumulator where register $\acc$ updates to $0$ when $\reset$ is $1$ (lines~2--3) and to $\acc + \iin$ otherwise (lines~4--5), whenever the driving clock signal \texttt{clock} transitions from $0$ to $1$, known as a positive clock edge (as specified by the \texttt{posedge} keyword in line~1).
To illustrate this dynamic behavior, the table in Figure~\ref{fig:background} shows an example of how variable values evolve over time:
\begin{itemize}
    \item
    The ``Cycle'' column enumerates clock cycles, where the number $i$ indicates that this row shows the values of all variables during the $i$-th \emph{clock cycle}, denoted $\Cycle_i$ hereafter.
    Here, a clock cycle is the time interval between consecutive positive edges of the clock signal \texttt{clock}.
\end{itemize}

\vspace{-1.1ex}

\begin{itemize}
    \item
    The ``Input'' columns provide the values of input variables such as $\reset$ and $\iin$, which are not defined within this code snippet but are externally supplied each cycle (e.g., by toggling a switch or pressing a button during circuit execution).
    \item
    The ``$\acc$'' column shows the value of register $\acc$ over time.
    For simplicity, we use $x_i$ to denote the value of variable $x$ during $\Cycle_i$.
    Initially, all registers hold undefined values at $\Cycle_0$, i.e., $\acc_0 = \uundefined$ in this example.
    During each cycle, lines 2--5 execute, and the \emph{new register value takes effect in the next cycle}, reflecting the special semantics of Verilog's non-blocking assignment operator \texttt{<=}, which models clock-driven updates of hardware registers---unlike software assignments.
    For example,
    (1) during $\Cycle_0$, $\acc\ \texttt{<=}\ 0$ (line 3) executes since $\reset_0 = 1$, making $\acc_1 = 0$;
    (2) during $\Cycle_1$, $\acc\ \texttt{<=}\ \acc + \iin$ (line 5) executes since $\reset_1 = 0$, giving $\acc_2 = \acc_1 + \iin_1 = 0 + 2 = 2$;
    (3) during $\Cycle_2$, the same update applies again, yielding $\acc_3 = \acc_2 + \iin_2 = 2 + 3 = 5$.
\end{itemize}
For further clarity, the lower portion of Figure~\ref{fig:background} presents a cycle-expanded control flow graph (CFG) that visualizes this dynamic behavior by highlighting execution paths in orange and blue.
In this representation, $\Entry_i$ denotes entry to the \texttt{always} block body (lines 2--5) during $\Cycle_i$.

\paragraph{Reset}

The \emph{reset signal} represents one of the most fundamental control signals critical for \emph{reliable} behavior in digital design~\cite{reset}, ensuring circuits begin operation from a known initial state rather than the default undefined state at power-up.
For example, in $\Cycle_0$ of Figure~\ref{fig:background} (highlighted in orange), $\reset_0$ is set to 1, ensuring that normal operation of the accumulator begins with a determined $\acc = 0$ instead of the undefined $\acc = \uundefined$.
We refer to the cycle in which the reset signal is active (i.e., $\reset = 1$) as the \emph{reset cycle}.
This reset cycle is essential for reliable normal operation: only after that cycle can we guarantee that during any subsequent non-reset $\Cycle_i$ (where $i \ge 1$ and $\reset_i = 0$), the accumulator \emph{reliably} maintains the invariant $\acc_{i+1} = \sum_{j = 1}^{i} \iin_j$.

By convention, any \emph{legal execution} of a circuit with reset capability should begin with a reset cycle to establish an initial state, followed by non-reset cycles for normal operation, as shown in Figure~\ref{fig:background}.
This creates a clear division of responsibility: hardware developers should ensure their Verilog programs function reliably under all legal executions, while hardware users should perform legal executions (i.e., reset before normal operation).
Consequently, the primary bugs of interest to developers are those that can occur during legal executions, as these fall within their expected domain of responsibility.

\begin{figure}[tp]
    \centering
    \begin{minipage}[b]{0.45\textwidth}
        \centering
        \begin{subfigure}[b]{\textwidth}
            \centering
            \includegraphics[trim=0cm 7.94cm 7.55cm 0cm,clip,page=3]{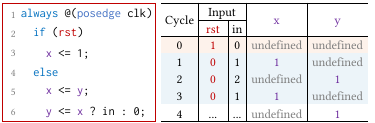}
            \caption{A simple missing-reset example.}\label{fig:missing-reset-eg}
        \end{subfigure}
        
        \vspace{0.3cm}
        
        \begin{subfigure}[b]{\textwidth}
            \centering
            \includegraphics[trim=0cm 5.04cm 7.55cm 2.17cm,clip,page=3]{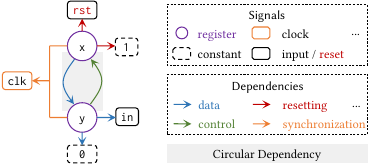}
            \caption{The hardware dependency graph of (\subref{fig:missing-reset-eg}).}\label{fig:hardware-dep}
        \end{subfigure}
    \end{minipage}%
    \hfill
    \begin{minipage}[b]{0.53\textwidth}
        \centering
        \begin{subfigure}[b]{\textwidth}
            \centering
            \includegraphics[width=\linewidth,trim=6.83cm 4.7cm 0 0,clip,page=3]{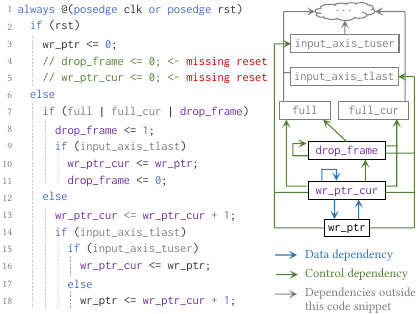}
            \caption{A real-world missing-reset example~\cite{asplos22bugbed} containing the first two bugs in our evaluation (Table~\ref{tab:bugs}).}\label{fig:missing-reset-real}
        \end{subfigure}
    \end{minipage}
    \caption{Missing-reset examples and their hardware dependency graphs.}
    \label{fig:missing-reset}
\end{figure}

\paragraph{Problem Formulation}\label{par:formulation}
The \emph{missing-reset} problem occurs on a register if there exists a \emph{legal execution} path (i.e., a reset cycle followed by non-reset cycles, as illustrated in Figure~\ref{fig:background}) in which this register's undefined value is \emph{infinitely often} used.
``Infinitely often'' is a standard term in mathematics~\cite{jacm1986temporallogic}.
In our context, it means that for a register $x$, after any given $\Cycle_N$, the undefined value of $x$ will be used again in some later cycle $\Cycle_n$ where $n > N$.
This problem is particularly dangerous because it implies that a register may unpredictably hold undefined values at any time, even during a legal execution, undermining hardware reliability and potentially leading to fatal functional flaws.
In contrast, finite occurrences of undefined values are common and acceptable in hardware (e.g., in pipelined designs~\cite{pipeline}), and such transient behavior should not be considered a bug.

Figure~\ref{fig:missing-reset-eg} illustrates a missing-reset example in which registers $x$ and $y$ both exhibit this problem, as shown in the table on the right.
In the reset cycle $\Cycle_0$ (shaded in orange), line~3 executes, updating $x$ from $\uundefined$ to $1$ in $\Cycle_1$, while $y$ remains $\uundefined$ because it is not assigned in this cycle.
In each non-reset cycle $\Cycle_i$ ($i \ge 1$, shaded in blue), lines~5--6 execute, resulting in the value updates $x_{i+1} = y_i$ and $y_{i+1} = x_i\ \mathtt{?}\ \mathtt{in}_i\ \mathtt{:}\ 0$, according to the Verilog non-blocking assignment semantics introduced in the \emph{Registers} paragraph.
More concretely, after $\Cycle_1$, we have $x_2 = y_1 = \uundefined$ and $y_2 = (\mite{x_1}{\mathtt{in}_1}{0}) = (\mite{1}{1}{0}) = 1$;
after $\Cycle_2$, we have $x_3 = y_2 = 1$ and $y_3 = (\mite{x_2}{\mathtt{in}_2}{0}) = (\mite{\uundefined}{2}{0}) = \uundefined$; and so on.
As a result, the undefined value propagates back and forth between $x$ and $y$, as reflected by the table in Figure~\ref{fig:missing-reset-eg}, demonstrating the missing-reset problem where the undefined values of $x$ and $y$ are used infinitely often.
To fix this issue, the developer should add a reset for $y$ in $\Cycle_0$, in addition to the existing reset for $x$ (line~3), ensuring that $y_1$ is defined and preventing the undefined values of $x$ and $y$ from being used infinitely often.

\subsubsection{Missing-Reset vs. UBI}\label{sec:motivation-vs}

The missing-reset problem in Verilog bears a resemblance to the classical \emph{use-before-initialization} (UBI) problem in software, which is prevalent in large-scale systems such as the Linux kernel and has driven decades of extensive research into static analyses for UBI detection~\cite{fse2020ubitect,oopsla2024enhance}.
Despite their shared focus on undefined-value-related bugs, analyses for missing resets are fundamentally distinguished from those for UBIs due to their \emph{hardware-specific nature}:\looseness=-1
\begin{itemize}
    \item
    \emph{Hardware-Specific Problem Formulation:}
    The missing-reset problem concerns a register whose undefined value is \emph{infinitely often} used in a \emph{legal} Verilog execution, whereas the UBI problem concerns a variable whose undefined value is \emph{once} used in \emph{any possible} software execution.
    Consequently, analyses for missing resets must establish hardware-specific insights to handle the ``infinitely often'' and ``legal'' constraints that do not arise in the UBI formulation.
    These insights will be elaborated in Section~\ref{sec:motivation-analysis}.
    \item
    \emph{Hardware-Specific Information Requirements:}
    The missing-reset problem is inherently grounded in hardware semantics, including clocks, registers, and resets, beyond the scope of UBI.
    Therefore, detecting such bugs naturally requires fundamental hardware information that goes beyond the needs of traditional UBI analyses.
    Notably, obtaining all sorts of fundamental information from Verilog programs demands a series of non-trivial hardware analyses, a challenge further elaborated in Section~\ref{sec:motivation-challenges}.
\end{itemize}
These distinctions highlight the necessity of establishing hardware-specific insights to design hardware bug detection analyses capable of capturing hardware-specific problems and fundamental hardware analyses that can retrieve essential hardware semantic information from Verilog programs.

\subsubsection{The Insight of Our Missing-Reset Analysis}\label{sec:motivation-analysis}

The key insight of our \aname{missing-reset} analysis is that only circuits involving \emph{circular dependencies} among registers can have missing-reset problems.
To see why, consider the opposite case:
if a circuit has no circular dependencies among registers, then a topological order must exist for these registers.
Following this order, defined values from constants or inputs can eventually propagate to all registers, ensuring that no register's undefined value is infinitely often used in any legal execution.\looseness=-1

Based on this hardware-specific insight, our \aname{missing-reset} analysis identifies registers that lack proper reset by reporting unreset registers involved in circular dependencies within the \emph{hardware dependency graph}.
This graph captures the data, control, resetting, and synchronization dependencies among hardware signals, including registers, clocks, resets, inputs, and constants.
For instance, Figure~\ref{fig:hardware-dep} shows the hardware dependency graph of the missing-reset example from Figure~\ref{fig:missing-reset-eg}.
We explain this hardware dependency graph as follows.
\begin{itemize}
    \item
    The blue edge from $x$ to $y$ represents that $x$ is data-dependent on $y$ (introduced by \texttt{x <= y} in line~5 of Figure~\ref{fig:missing-reset-eg}), while the green edge from $y$ to $x$ indicates that $y$ is control-dependent on $x$ (introduced by \texttt{y <= x ? ...} in line~6 of Figure~\ref{fig:missing-reset-eg}).
    Together, these edges form a circular dependency (highlighted in the gray region) between $x$ and $y$, underlying the situation where undefined values propagate back and forth between them, as elaborated in the \emph{Problem Formulation} paragraph of Section~\ref{sec:motivation-problem}.
    \item
    The orange synchronization edges from $x$ and $y$ to the clock signal $\mathtt{clk}$ indicate that $x$ and $y$ are registers synchronized by the clock $\mathtt{clk}$.
    \item
    The red resetting edges from $x$ to the reset signal $\mathtt{rst}$ and the constant $1$ capture $x$'s reset logic---resetting $x$ to $1$ under the control of $\mathtt{rst}$.
\end{itemize}

More concretely, \aname{missing-reset} includes the following four steps:
(1)~\emph{Non-Synthesizable Code Exclusion}:
Verilog codebases typically include non-synthesizable portions written for simulation purposes, which do not appear in the final synthesized hardware and cannot contain real hardware bugs.
Thus, we exclude such code from our analysis.
(2)~\emph{Hardware Dependency Graph Construction}:
Build the hardware dependency graph by resolving all sorts of dependencies in the Verilog program (e.g., as shown in Figure~\ref{fig:hardware-dep}).
(3)~\emph{Circular Dependency Identification}:
Locate cycles in the hardware dependency graph (the gray region in Figure~\ref{fig:hardware-dep}) to identify registers involved in circular dependencies (e.g., $x$ and $y$ in Figure~\ref{fig:hardware-dep}).
(4)~\emph{Missing Reset Reporting}:
Report registers from step~(2) that lack proper reset.
In Figure~\ref{fig:hardware-dep}, $y$ is reported because it is unreset, whereas $x$ is not since it is reset to $1$ by $\mathtt{rst}$ (indicated by the red resetting edges in Figure~\ref{fig:hardware-dep}).

To provide a glimpse into real-world missing-reset bugs, Figure~\ref{fig:missing-reset-real} presents such an example along with its hardware dependency graph, which retains only data and control dependencies for clarity.
The code snippet originates from a FIFO (first-in, first-out) buffer in the Verilog-AXIS project~\cite{verilog-axis}, a popular open-source design with over 800~GitHub stars.
Our analysis detected two missing-reset bugs in the registers \texttt{drop\_frame} and \texttt{wr\_ptr\_cur} (lines~4--5 in Figure~\ref{fig:missing-reset-real}) through the four steps described above.
These bugs are \emph{fatal}: the missing reset of \texttt{drop\_frame}, a frame-dropping state register, leads to unintended packet drops, while the missing reset of \texttt{wr\_ptr\_cur}, a write-pointer register, causes writes to incorrect buffer addresses.

\subsubsection{The Challenge in Building Sophisticated Hardware Static Analysis}\label{sec:motivation-challenges}

Even with the core insights established, building a sophisticated hardware analysis remains challenging due to the lack of analyses focused on extracting fundamental hardware information---a topic that few prior works have explored.

For example, to detect missing-reset problems involving hardware \emph{registers} driven by \emph{clock} signals and reset by \emph{reset} signals, as guided by the insights established in Section~\ref{sec:motivation-analysis}, the analysis requires essential hardware information about whether each variable represents a hardware register, a clock signal, a reset signal, or none of these.
However, this is nontrivial because Verilog variables do not explicitly declare such information.
This information is typically revealed after lengthy synthesis processes that map Verilog code to physical hardware components based on behavioral semantics.
To bridge this gap for both the \aname{missing-reset} analysis and other clients requiring this hardware information, we developed hardware understanding analyses that reason about such behavioral semantics (e.g., inferring registers, clocks, and reset signals), enabling us to efficiently obtain the necessary information for \aname{missing-reset} analysis without resorting to heavyweight synthesis.\looseness=-1

Moreover, these hardware understanding analyses require more fundamental hardware information, which is also crucial for bug detection clients and future potential clients, such as hardware security analyses. This necessitates diverse analyses, including but not limited to:
(1) Verilog module hierarchy analysis for inter-module reasoning; 
(2) hardware data-flow and control-flow analyses for constructing various graph structures (e.g., the hardware dependency graph used by the \aname{missing-reset} analysis);
(3) analyses for capturing hardware's synchronization and concurrency characteristics, which are essential for reasoning about behavioral semantics in hardware understanding;
(4) analyses of bit selects/part selects and bit-vector arithmetic, which are indispensable for achieving practically useful precision in hardware analysis.

Ultimately, building sophisticated hardware static analysis challenges us to uncover diverse fundamental analyses, understand their interdependencies, and enable their interoperability.
To this end, we introduce the well-organized analysis suite in Figure~\ref{fig:analyses}.
Next, we'll discuss the analyses that support the \aname{missing-reset} analysis.

\subsection{Supporting Sophisticated Analyses: A Case Study}\label{sec:supporting-case-study}

\begin{wrapfigure}[19]{r}{0.4\textwidth}
    \centering
    \vspace{-2.5ex}
    \includegraphics[width=\linewidth]{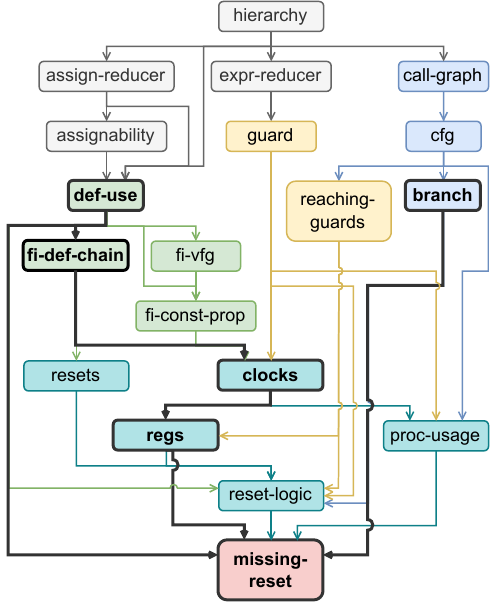}
    \vspace{-4ex}
    \caption{All analyses that support the \aname{missing-reset} analysis.}
    \label{fig:missing-reset-dep}
\end{wrapfigure}

Although the four steps of \aname{missing-reset} analysis are succinctly described in Section~\ref{sec:motivation-analysis}, their implementation relies on the collaboration of nearly twenty analyses in our suite, totaling 9,700 lines of code.
Figure~\ref{fig:missing-reset-dep} illustrates all analyses supporting \aname{missing-reset}, arranged in topological order from fundamental analyses (top) to analysis clients (bottom).
Below, we use Figure~\ref{fig:missing-reset-dep} to help readers understand why the \aname{missing-reset} analysis requires its dependencies and how these dependencies contribute to its overall effectiveness, thereby providing insight into the organization of our analysis suite for supporting sophisticated analyses.
For readability, we do not discuss algorithmic details here; interested readers can find them in our open-source project.

\paragraph{Interdependencies}
To clarify why the \aname{missing-reset} analysis requires its dependencies in Figure~\ref{fig:missing-reset-dep} while maintaining simplicity and readability, we backtrack the bold dependency arrows in the figure, emphasizing both the functionality of these analyses and the necessity of their interdependencies.
To begin,
(1)~Since missing-reset bugs are confined to registers, \aname{missing-reset} directly depends on the \aname{regs} analysis to infer physical registers, and utilizes \aname{def-use} and \aname{branch} analyses to build a graph containing data and control dependencies for cycle detection, as described in Section~\ref{sec:motivation-analysis};
(2)~Since registers are synchronized by clock signals, \aname{regs} tightly depends on \aname{clocks} to infer physical clocks;
(3)~Since clocks are propagated as data across modules in Verilog, the \aname{clocks} analysis directly depends on \aname{fi-def-chain} (where \texttt{fi} denotes flow-insensitive analysis), which provides an inter-module graph representing data flow to support such reasoning;
(4)~Lastly, \aname{fi-def-chain} leverages the \aname{def-use} analysis to obtain define-use relations between variables, including inter-module relations, to build the inter-module graph.
The lighter dependency arrows in Figure~\ref{fig:missing-reset-dep} also support \aname{missing-reset} in a similar way to the bold ones; for brevity, we omit a redundant discussion of these dependencies.

\paragraph{Analysis and Its Downstream Effectiveness}
These interdependencies underscore that the effectiveness of each analysis can be critical for downstream analyses.
For example, reduced precision in the \aname{clocks} analysis (i.e., misclassifying non-clock variables as clocks) directly affects the precision of \aname{regs}, since registers are synchronized by clock signals; insufficient precision in \aname{regs} can lead to excessive false reports in \aname{missing-reset}, undermining its practical value.
Likewise, insufficient recall in \aname{clocks} and \aname{regs} (i.e., reduced soundness that omits some true behaviors) may cause \aname{missing-reset} to overlook genuine missing-reset problems.
To address these challenges, key supporting analyses are designed to achieve high effectiveness.
For readability, we present our design considerations for two representative analyses: \aname{regs} and \aname{fi-const-prop}.
The former shows our approach to designing a hardware analysis, which is not found in software, to achieve high recall and precision.
The latter—a flow-insensitive constant propagation analysis—shows that some analyses in our suite, though inspired by software concepts, require Verilog-specific adaptations to achieve effective results on hardware.

\aname{regs} (the register inference analysis) predicts which Verilog variables are synthesized as hardware registers and identifies their driven clocks.
To achieve high precision, it cannot simply report every variable declared with Verilog's \texttt{reg} keyword, because many such declarations denote intermediate variables rather than physical registers (the keyword is misleading), yielding numerous false positives.
Instead, \aname{regs} classifies registers by their semantics.
Register updates are synchronized to clock edges, which are expressed in Verilog as non-blocking assignments guarded by event controls that reference clock signals;
accordingly, the core step of \aname{regs} is to locate such non-blocking assignments and report their left-hand-side (LHS) variables as registers.
Recall that Figure~\ref{fig:background} illustrates an event control (e.g., \lstinline[language=Verilog,style=vstyle]|@(posedge clock)|) guarding a non-blocking assignment (e.g., \lstinline[language=Verilog,style=vstyle]|acc <= acc + in|), where the LHS variable \texttt{acc} is the register.
To achieve high recall, the key is to soundly extract the guarding relationship between non-blocking assignments and event controls.
Simply matching the pattern shown in Figure~\ref{fig:background} provides unsound guarding-relationship results because event controls and non-blocking assignments can appear in arbitrary orders within an \texttt{always} block, yielding nontrivial guarding relationships.
For example, the body of an \texttt{always} block could be \lstinline[language=Verilog,style=vstyle]|@(posedge clock1); x <= 1; @(negedge clock2); y <= 0;|, where \lstinline[language=Verilog,style=vstyle]|y <= 0| is guarded by the later \lstinline[language=Verilog,style=vstyle]|@(negedge clock2)|, not the earlier \lstinline[language=Verilog,style=vstyle]|@(posedge clock1)|.
Thus, we compute the guard relation via a sound data-flow analysis over the control-flow graph using the fundamental \aname{reaching-guards} analysis, on which \aname{regs} depends, as shown in Figure~\ref{fig:missing-reset-dep}.

\aname{fi-const-prop} is widely used in our suite, and its precision is crucial for the effectiveness of downstream analyses.
Unlike typical software programs, achieving high precision in Verilog requires bit-level constant reasoning, as Verilog programs frequently operate on individual bits or bit-vectors corresponding to physical wires.
Accordingly, \aname{fi-const-prop} accurately models Verilog's four-valued logic (0/1/X/Z) and bit-vector arithmetic to enable precise bit-level constant propagation.
We discuss a bug detected thanks to this design in Section~\ref{sec:bug}.
Note that we do not adopt the flow-sensitive variant of constant propagation commonly used in software analysis, as we found it provides only marginal precision improvement while notably reducing analysis speed.
This is mainly because most Verilog variables are defined in a single location, reflecting the fact that hardware signals typically have only one definition, which makes flow-sensitivity less beneficial.
This further underscores the necessity of Verilog-specific adaptation.

\paragraph{Discussion on Soundness}
As mentioned above, unsoundness in our hardware understanding analyses (e.g., \aname{clocks} and \aname{regs}) can render \aname{missing-reset} unsound. 
Although we strive to maintain their high recall in practice, it is important to note that perfect soundness for these analyses is inherently unattainable.
This arises because these hardware understanding analyses attempt to predict the circuit structure produced by synthesis tools, while the Verilog standard~\cite{verilog-std-2005} specifies only circuit behaviors, not their structure.
Consequently, different synthesis tools may generate structurally distinct circuits from the same Verilog code, with no formal standard serving as ground truth for sound prediction.
Nevertheless, in practice, synthesis tools tend to follow common mapping conventions, allowing our analyses to effectively predict useful synthesis outcomes despite this unavoidable limitation.

Excluding this source of unsoundness, the \emph{reachable closure} of \aname{missing-reset}'s results---the set of registers reachable from the reported registers in the hardware dependency graph---can \emph{soundly} capture all possible missing-reset problems.
Registers outside this closure cannot exhibit missing-reset issues, as they do not participate in circular dependencies within the circuit.
Based on our insight in Section~\ref{sec:motivation-analysis}, their values are ultimately computed from constants, inputs, or reset registers, all of which obtain defined values after the reset cycle in any legal execution.

\vspace{1ex}
The collaborative nature of our analyses goes beyond those mentioned above and can be more intricate when developing hardware static analyses that are more sophisticated.
We hope the example of \aname{missing-reset} can serve as a reference for developers to create their own applications by leveraging existing analyses in our analysis suite.

\section{Analysis-Oriented IR and Front End} \label{sec:ir}

The design of the IR and front end, a classic topic in compilers, requires further care in the context of a hardware analysis framework.

\begin{figure}[p]
\centering
\footnotesize
\input{figures/ir-syntax}
\caption[Syntax of Qihe IR]{The syntax of Qihe IR, which uses \mterm{fixed-width} \mterm{bold} fonts for terminals, $\msym{italics}$ for non-terminals, and \mcomment{gray} fonts to omit common definitions for brevity.
The symbol $\msym{pattern}^?$ means $\msym{pattern}$ is optional. $\msym{pattern}^*$ means zero or more occurrences of $\msym{pattern}$.
Section~\ref{sec:ir} provides a detailed example (Figure~\ref{fig:ir}) to highlight the key design features of Qihe IR.}
\label{fig:ir-syntax}
\end{figure}

\begin{figure}[t]
\centering
\includegraphics[width=\linewidth,trim={0cm 7.5cm 0cm 0cm},clip]{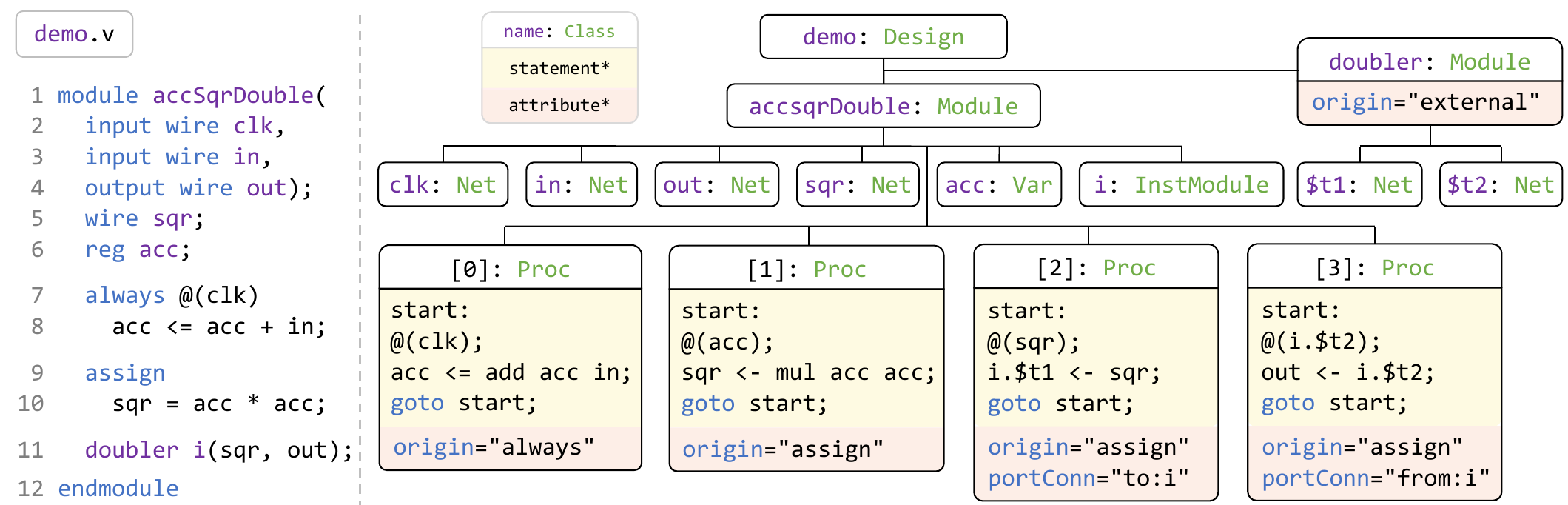}
\caption{
Example of Qihe IR derived from a Verilog program.
Each square on the right represents an IR node modeling Verilog's hierarchical language construct on the left, as elaborated in Section~\ref{sec:ir-hierarchy}.
Specific nodes (e.g., \texttt{Proc}) contain three-address code statements, detailed in Section~\ref{sec:ir-3ac}.
Each node can have zero or more attributes, whose uses are discussed in Sections~\ref{sec:attr} and~\ref{sec:incomplete}.
Nodes with \texttt{\$}-prefixed names are generated by the front end.
}
\label{fig:ir}
\end{figure}

Regarding the IR, Qihe introduces a hardware program abstraction called Qihe IR, which is designed to capture all Verilog features for describing hardware\footnote{The Verilog language includes synthesizable constructs for describing hardware and non-synthesizable ones for writing testbenches. Non-synthesizable features are not relevant to Qihe's hardware analysis and are thus excluded from this paper. When referring to Verilog in this context, we focus on its synthesizable subset. However, Qihe does support certain commonly used non-synthesizable features to expand its application scope. Additionally, Qihe also supports frequently used features from SystemVerilog~\cite{systemverilog-std-2017}, an extension of Verilog.} while remaining simple enough to facilitate the development of analyses.
Figure~\ref{fig:ir-syntax} presents the syntax of Qihe IR.
Apart from some basic definitions, Qihe IR consists of two main components: hierarchical structures and three-address code (3AC).
Instead of elaborating on the full syntax and semantics of Qihe IR, which would be verbose and difficult to follow, we use an example in Figure~\ref{fig:ir} to highlight the key designs of these two components in Sections~\ref{sec:ir-hierarchy} and \ref{sec:ir-3ac}, respectively.
Readers can refer to our project documentation for additional details about Qihe IR.

Regarding the front end, beyond its standard task of converting Verilog code into Qihe IR, we find that it must also cooperate with Qihe IR to address specific needs in hardware analysis.
We highlight two important aspects:
(1) how to satisfy the seemingly conflicting requirements of varied analyses (Section~\ref{sec:attr}), and 
(2) how to handle incomplete programs to enable sophisticated analyses (Section~\ref{sec:incomplete}). 
Additional design and implementation details of the front end are available in Qihe's open-source repository.

\subsection{Intuitive Tree Representation for Verilog's Hierarchical Structure} \label{sec:ir-hierarchy}

A Verilog design consists of modules that can include nets, variables, sub-module instantiations, functions, and other constructs, forming a nested hierarchy.
Qihe IR's high-level structure is designed to closely mirror this hierarchical organization of hardware designs.
As illustrated in the \emph{Hierarchical Structures} portion of Figure~\ref{fig:ir}, the Qihe IR syntax begins with a design that contains modules.
These modules, in turn, include nets, variables, module instantiations, process blocks, and functions, directly reflecting Verilog's hierarchical structure.
This design enables users familiar with Verilog to quickly understand the IR and navigate it intuitively using only the Verilog source code as a reference when developing analyses.

In this section, we use the example in Figure~\ref{fig:ir} to illustrate how Qihe IR's hierarchical structure corresponds to Verilog's structure, explaining the semantics of Verilog constructs along the way for readers who may be less familiar with Verilog.
This example covers the most commonly used constructs in Verilog.

As illustrated in Figure~\ref{fig:ir}, the left side shows a Verilog program named \texttt{demo.v}, while the right side presents its corresponding Qihe IR.
The hierarchical structure of Qihe IR is illustrated as a tree rather than in textual syntax, because the tree representation more intuitively conveys the hierarchical relationships among different Verilog constructs and is closer to the in-memory representation of Qihe IR used by analyses.
Each node in the tree corresponds to an IR construct and is labeled with a name and a class.
The name identifies the node and is derived from the Verilog code when possible.
The class indicates which Qihe IR construct the node represents, corresponding to the non-terminals defined in Figure~\ref{fig:ir-syntax}. 
Let's examine the tree from top to bottom.

The root of the IR is a node named \texttt{demo} of class \mclass{Design}, representing the entire hardware design described by the Verilog code in \texttt{demo.v} on the left of Figure~\ref{fig:ir}.
The name \texttt{demo} is derived from the Verilog file name \texttt{demo.v}.
The \mclass{Design} class corresponds to the $\msym{design}$ non-terminal in Figure~\ref{fig:ir-syntax}.
We omit detailed explanations of node names and classes in subsequent discussions unless necessary, as they are largely self-explanatory.

The root node has two \mclass{Module} nodes as its children, each representing a Verilog module, reflecting the fact that a Verilog design is composed of a collection of module definitions.
In Verilog, a module is a fundamental unit for encapsulating reusable circuit designs, similar to how a function encapsulates reusable computation logic, and can have inputs and outputs known as ports.
For example, the left part of Figure~\ref{fig:ir} defines a Verilog module named \texttt{accSqrDouble} (line~1) with inputs \texttt{clk} and \texttt{in} (lines~2--3) and an output \texttt{out} (line~4).
This module encapsulates a circuit that accumulates input values from \texttt{in} into a register \texttt{acc} synchronized by the clock \texttt{clk} (lines~7--8), squares (lines~9--10) and doubles (line~11) the value from \texttt{acc}, and outputs the result to \texttt{out} (line~11).
Let's follow the \texttt{accSqrDouble} node to elaborate on these details.

The \texttt{accSqrDouble} node has several subnodes representing different constructs within the module that describe the circuit.
To begin with, \mclass{Net} and \mclass{Var} correspond to Verilog's \texttt{wire} (lines~2--5) and \texttt{reg} (line~6), respectively.
These constructs are used to describe physical registers and wires.
Note that module ports are special \texttt{wire}s or \texttt{reg}s with additional \texttt{input} or \texttt{output} keywords (lines~2--4).
Although Verilog includes additional keywords like \texttt{wand} and \texttt{wor} for declaring special types of wires, Qihe IR uniformly represents all wire-related constructs using the \mclass{Net} class for simplicity.
Furthermore, Qihe IR adopts the term \mclass{Var} instead of \mclass{Reg} to represent Verilog's \texttt{reg} declarations.
This is because, in Verilog, the \texttt{reg} keyword does not necessarily imply a physical register in hardware; rather, it simply declares a software-like variable that may be mapped to wires, registers, or optimized out entirely based on its usage in the design.
To know whether a \mclass{Var} is mapped to a physical register, developers can utilize Qihe's \aname{reg} analysis, which provides this information, as discussed in Section~\ref{sec:supporting-case-study}.

\begin{wrapfigure}[9]{r}{4cm}
\vspace{-3.5ex}
\begin{lstlisting}[language=verilog,numbers=none,style=vstyle]
module doubler(
  input wire $t1,
  output wire $t2);
  // double $t1 and
  // assign to $t2
endmodule
\end{lstlisting}
\vspace{-2ex}
\caption{The inferred signature of module \texttt{doubler} in Figure~\ref{fig:ir}.}\label{fig:infer}
\end{wrapfigure}
Next, the \texttt{accSqrDouble} node includes a subnode of class \mclass{InstModule} due to the module instantiation on line~11.
Module instantiation is a fundamental mechanism in Verilog for reusing encapsulated circuit logic.
In this example, the line \texttt{doubler i(sqr, out)} first creates an instance named \texttt{i} of the module \texttt{doubler}.
The definition of \texttt{doubler} resembles the one shown in Figure~\ref{fig:infer}, which encapsulates a circuit that doubles its input value and outputs the result.
The instantiation next connects the wire \texttt{sqr} to the input port \texttt{\$t1} of the module instance \texttt{i} and the wire \texttt{out} to the output port \texttt{\$t2}, ensuring that \texttt{out} carries twice the value of \texttt{sqr} without requiring the doubling logic to be redefined.
It is important to note that the \mclass{InstModule} node represents only the module instance \texttt{i}, while the port connections are modeled separately by \mclass{Proc} nodes. This design choice is further elaborated in the following paragraph.
Additionally, the definition of a module like \texttt{doubler} may be unavailable in many real-world scenarios due to the widespread use of third-party IP cores with intellectual property restrictions.
To handle such cases, our front end automatically infers key information about the missing module, such as the types and directions of its ports.
The code shown in Figure~\ref{fig:infer} is actually inferred by our front end when processing the Verilog program in Figure~\ref{fig:ir} without the definition of \texttt{doubler}, and the top-right corner of Figure~\ref{fig:ir} illustrates a subtree representing this inferred module.
This inferred information is critical for enabling sophisticated analyses, and we elaborate on the inference process in Section~\ref{sec:incomplete}.

Finally, the \texttt{accSqrDouble} node includes four \mclass{Proc} nodes that unify the representation of Verilog \texttt{always} blocks (lines~7--8), continuous assignments (lines 9--10), and port connections (line~11).
These three constructs are among the most commonly used for describing circuit logic in Verilog.
However, they differ significantly in syntax, which complicates the development of analysis logic.
Without unification, developers must handle each construct separately, resulting in redundant and complex code.
To address this challenge, Qihe IR introduces the \mclass{Proc} node (process blocks), which uses three-address code (3AC) to provide a uniform representation of these constructs.
This unification is made possible because, under Verilog's simulation semantics~\cite{lambdav,verilog-std-2005}, all these constructs can be interpreted as concurrent processes that execute statements to emulate the behavior of physical circuits.
The term \mclass{Proc} is used to reflect this.
In Section~\ref{sec:ir-3ac}, we will elaborate on how these constructs describe circuit logic and how 3AC is designed to model them uniformly.

\subsection{Unified Three-Address Code Representation for Verilog's Circuit Logic}\label{sec:ir-3ac}

In addition to Verilog's hierarchical structure, Qihe IR uses three-address code (3AC) to model circuit logic.
The 3AC design simplifies the development of analyses by representing various Verilog constructs in a straightforward and uniform manner. 
Figure~\ref{fig:ir-syntax} illustrates all the 3AC statements in Qihe IR.
For readers familiar with Verilog, many of these statements intuitively correspond to Verilog constructs.
As a result, explaining each 3AC statement in detail in this paper would be redundant and tedious, so we have left these specifics to our documentation.
Instead, we focus on a key design feature: how 3AC is used to uniformly model different Verilog constructs for describing circuit logic.
Using the example in Figure~\ref{fig:ir}, we explain the semantics of the three most frequently used Verilog constructs: \texttt{always} blocks, continuous assignments, and port connections, and how the circuit logic described by them is modeled using 3AC in Qihe IR.

\paragraph{Always Blocks}
An \texttt{always} block uses statements to specify software-like computation logic between variables, which the Verilog synthesizer maps to circuit logic between physical wires and registers.
In this example, the body of the \texttt{always} block contains the statement \texttt{@(clk) acc <= acc + in}, meaning that whenever \texttt{clk} changes (\texttt{@clk}), the register \texttt{acc} is updated by adding \texttt{in} to its previous value.
Note that Verilog introduces a special non-blocking assignment operator, \texttt{<=}, which is unique and essential for distinguishing between register updates and software-like variable assignments within \texttt{always} blocks.
The behavior of the \texttt{always} block is modeled by the 3AC statements within the \texttt{[0]:Proc} node\footnote{Since \texttt{always} blocks, continuous assignments, and port connections are unnamed, we use indices (\texttt{[0]}, \texttt{[1]}, etc.) to refer to them by their sequence of occurrence in the source code.}.
\texttt{[0]:Proc} contains an infinite loop that monitors the clock via a statement \texttt{@(clk)} and updates \texttt{acc} using \texttt{add acc in} whenever \texttt{clk} changes, directly corresponding to the behavior of the \texttt{always} block.
To correctly model register updates in Verilog, we retain the non-blocking assignment operator \texttt{<=} as in line~8.

\paragraph{Continuous Assignments}
Although \texttt{always} blocks are enough to describe any circuit logic, Verilog also provides continuous assignments as a simpler way to express combinational (i.e., stateless) logic.
A continuous assignment establishes a constraint between the wires and variables on the left-hand side and right-hand side of the assignment.
For example, \texttt{assign sqr = acc * acc} specifies that \texttt{sqr} should always hold the value of \texttt{acc * acc}.
The Verilog synthesizer generates the corresponding circuit logic for this multiplication to enforce this constraint.
This behavior is modeled by \texttt{[1]:Proc}, which monitors changes in the value of \texttt{acc} and updates \texttt{sqr} accordingly.
It resembles \texttt{[0]:Proc}, as continuous assignments can often be replaced by an equivalent \texttt{always} block; therefore, we unify them under \texttt{Proc}.
However, the key difference is the use of the \texttt{<-} assignment operator, which is designed in Qihe IR to support Verilog's multiple drivers feature.
Because of this feature, continuous assignments cannot always be fully replaced by \texttt{always} blocks.
In Verilog, it is legal for multiple continuous assignments to target the same wire, such as ``\texttt{assign u = 1; assign u = 0}''.
The final value of \texttt{u} is determined by merging these assigned values according to predefined rules.
For example, if \texttt{u} is declared as a \texttt{wire}, the merged value is Verilog's special unknown value \veriX;
if \texttt{u} is declared as a \texttt{wand} (wired-AND), the merged value is \texttt{0 \& 1 = 0}.
To model this behavior, we introduce the \texttt{<-} operator, which records all assigned values to \texttt{u}.
When there are multiple \texttt{<-} assignments, all recorded values are merged to determine the final value of \texttt{u}.

\paragraph{Port Connections}
Port connections are implicitly established during module instantiation and share the same semantics as continuous assignments.
For example, the instantiation of module \texttt{doubler} on line~11 of Figure~\ref{fig:ir} creates two port connections: one from \texttt{sqr} to \texttt{\$t1} of instance \texttt{i} and another from \texttt{\$t2} of instance \texttt{i} to \texttt{out}.
This means that whenever \texttt{sqr} changes, its value is assigned to \texttt{\$t1}, and whenever \texttt{\$t2} changes, its value is assigned to \texttt{out}.
This behavior is modeled by \texttt{[2]:Proc} and \texttt{[3]:Proc}, in a manner similar to how \texttt{[1]:Proc} models the continuous assignment.
Note that we use the hierarchical reference \texttt{i.\$t1} to refer to the \texttt{\$t1} port of instance \texttt{i} without ambiguity in the statement; the same applies to \texttt{i.\$t2}.

\vspace{1ex}
In summary, while Verilog provides convenient constructs that simplify hardware design, these features increase the complexity of circuit logic analysis.
By converting such constructs into Qihe's process blocks with 3AC statements, Qihe enables analysis developers to handle them uniformly.
During this conversion, Qihe IR also introduces statements and operators uncommon in software analysis IRs, such as guard statements and the \texttt{<-} operator, to accurately capture Verilog semantics.

The remaining part of Figure~\ref{fig:ir} not yet discussed is the attribute system, which enables key-value pairs to be associated with Qihe IR's hierarchical structures and statements. 
This system works closely with our front end to support additional analysis-oriented features, including those described in the following two sections.

\subsection{Extensible IR for Varied Analysis Requirements} \label{sec:attr}
In addressing the varied requirements of different analyses, we identified seemingly conflicting requirements:
(1) Some analyses require Qihe to simplify programs by reducing complex language constructs to their basic forms, relieving them from handling redundant details in the source code.
(2) Others require Qihe to preserve extensive source-level information, which might otherwise be lost during simplification, to help them infer user intentions or speed up analysis.

To satisfy both needs, Qihe IR is designed around a simple core augmented with an extensible attribute system:
(1) Its basic constructs form a small core of Verilog that is semantically equivalent to the full language, as discussed in previous sections.
This core simplifies Verilog by desugaring syntactic sugar and complex constructs into a uniform, simple representation.
As a result, developing analyses on this core becomes much simpler, as it avoids dealing with unnecessary source-level details and enables handling different Verilog constructs using the same IR constructs.
(2) Qihe IR augments the core with an attribute system so that the front end can supplement additional information about the original syntax patterns desugared during IR generation.
This allows specialized analyses to leverage these attributes, enhancing precision or improving performance.

For example, Verilog's \texttt{always} blocks, port connections, and continuous assignments are all converted to process blocks in Qihe IR, allowing analyses to handle them uniformly with ease.
Meanwhile, the attribute system preserves information about their original forms in the source code.
Consider the port connection from \texttt{sqr} to \texttt{i.\$t1} in Figure~\ref{fig:ir} as an example.
The connection is first desugared into a continuous assignment and then converted to a process block \texttt{[2]:Proc}.
The block has the attribute \texttt{portConn="to:i"}, which is added by the front end during desugaring, indicating that this is a port connection directed into the instance \texttt{i}.
It also has the attribute \texttt{origin="assign"}, which is added during the conversion to a process block, indicating that it originated from a continuous assignment (itself desugared from the port connection).
These attributes can guide the optimization of certain analyses.
For example, the \aname{fast-reaching-defs} analysis in Figure~\ref{fig:analyses} utilizes this information to achieve an average $1.7\times$ speedup across the 13 programs in Table~\ref{tab:analyses-results} (reducing runtime from 44.6\,s to 25.6\,s on the largest XS program with 1.8M lines).

\subsection{Automatic Signature Inference for Incomplete Programs} \label{sec:incomplete}
Incomplete programs are common in hardware design due to the widespread use of IP (intellectual property) cores whose implementations are encrypted or inaccessible.
While linters can ignore the missing modules and still perform local pattern-based analysis, to support sophisticated inter-module analyses with better precision and soundness, Qihe needs to properly handle the missing pieces, which typically requires the signatures of the missing modules.

A module's signature includes its name, port types and directions, plus metadata such as compile-time (elaboration-time) parameters.
Many of Qihe's more sophisticated analyses rely on this information.
For example, taint analysis must track how values propagate within or across modules and determine whether a value is tainted.
If a module lacks a signature---specifically, the direction of each port (input or output)---the taint analysis cannot propagate values at the module's ports because it does not know whether the value should flow into or out of the module, unless additional work is performed to infer this signature.
However, with our front end automatically inferring this signature, not only taint analysis but also other analyses involving inter-module flow can benefit significantly.
Another example is bit-width analysis. 
It requires the type and metadata information to check implicit type conversions that typically lead to bugs in Verilog.
When there are multiple instantiations of a missing module, if a port is connected to wires of different types in different instantiations (e.g., one 8-bit and one 16-bit) and no metadata appears to explain the mismatch, analyses should raise a warning about the implicit type conversion.

Qihe's front end is thus designed to automatically infer the missing module signatures from the available context properly, relieving analysis developers from having to do this repeatedly when implementing their own analyses.
We illustrate how Qihe handles a program with a missing module definition.

Consider the undefined module \texttt{doubler} in Figure~\ref{fig:ir}: when processing the module instantiation ``\texttt{doubler i(sqr, out)}'', the front end detects that the definition of the instantiated module \texttt{doubler} is absent.
To proceed, Qihe infers the signature of the module \texttt{doubler}, which includes the type and direction of its ports (this module appears to have no metadata).
(1) First, it determines the number of ports.
In this example, \texttt{doubler} has two ports, named \texttt{\$t1} and \texttt{\$t2}, corresponding to the two wires \texttt{sqr} and \texttt{out} connected to it.
(2) Next, Qihe infers the port type based on the type of the connected wires.
For instance, since \texttt{out} (connected to \texttt{\$t2}) has a one-bit type, Qihe assumes \texttt{\$t2} also has a one-bit type.
Here, the \texttt{doubler} is only instantiated once; however, if a module is instantiated multiple times, for the same port, we need to check all the wires connected to it in different instantiations.
If these wires share the same type, that type is used as the result; otherwise, their types are merged to determine the inferred result.
(3) Finally, the port direction is inferred by examining the definitions of the connected wires.
If a wire connected to the port is defined by other assignments, the port is unlikely to be an output port, as this would cause conflicts.
Thus, it is considered an input port.
Otherwise, it is treated as an output port.
In this example, \texttt{\$t1} is inferred as an input port because the wire \texttt{sqr} connected to it is already defined by lines~9--10, indicating that values flow from \texttt{sqr} into the instantiated module through \texttt{\$t1}.
In contrast, \texttt{\$t2} is inferred as an output port because the wire \texttt{out} connected to it is an output of the enclosing module (line~4) and is not otherwise driven, indicating that values are supplied by the instantiated module via \texttt{\$t2}.

Note that because the generated \texttt{doubler} module cannot be distinguished from a deliberately designed two-port module with an empty implementation, Qihe marks it as \texttt{external} using the attribute system, as shown in Figure~\ref{fig:ir}.
This guides subsequent analyses to make sound assumptions about its unknown implementation.
\section{Analysis Management}\label{sec:manage}

While traditional static analysis frameworks for software like Soot~\cite{soot} and WALA~\cite{wala} have gained wide adoption, they primarily focus on implementing standalone analyses rather than offering mechanisms for managing interdependent analyses.
These frameworks provide limited support for use cases like (1) easy integration of new analyses, and (2) automatic execution of analyses with intricate dependencies.

We argue that effective management of analyses, particularly the integration and execution of interdependent analyses, is crucial for establishing Qihe as a practical framework.
Without such capabilities, maintaining over 40 interdependent analyses, as illustrated in Figure~\ref{fig:analyses}, would become a challenging task.
The recent software static analysis framework Tai-e~\cite{taie} has also emphasized the importance of developer-friendly static analysis frameworks, introducing a configuration-file-based mechanism for analysis management, which received positive feedback from its users.
 
Inspired by Tai-e, we introduce a systematic mechanism for analysis management in Qihe, but with a different design.
Instead of requiring developers to write configuration files for analysis management, as in Tai-e, we provide Java annotations to achieve the same functionality (as Qihe is implemented in Java).
This approach simplifies the development of new analyses by utilizing Java annotations to automate tasks such as analysis discovery, which would otherwise require explicit registration in configuration files.
Moreover, it helps developers avoid careless mistakes by taking advantage of the mature Java type system and associated tools, such as code completion and compile-time validation, which are typically unavailable when using configuration files.

This section highlights the key design aspects of Qihe's analysis manager in terms of analysis integration and execution.
Beyond these core facilities, Qihe also provides additional management capabilities, such as analysis options and diagnostics handling.
For readers interested in further details, we encourage consulting our project documentation.

\begin{figure}[tp]
\centering
\includegraphics[width=\linewidth,trim={0cm 4.9cm 3.2cm 0cm},clip]{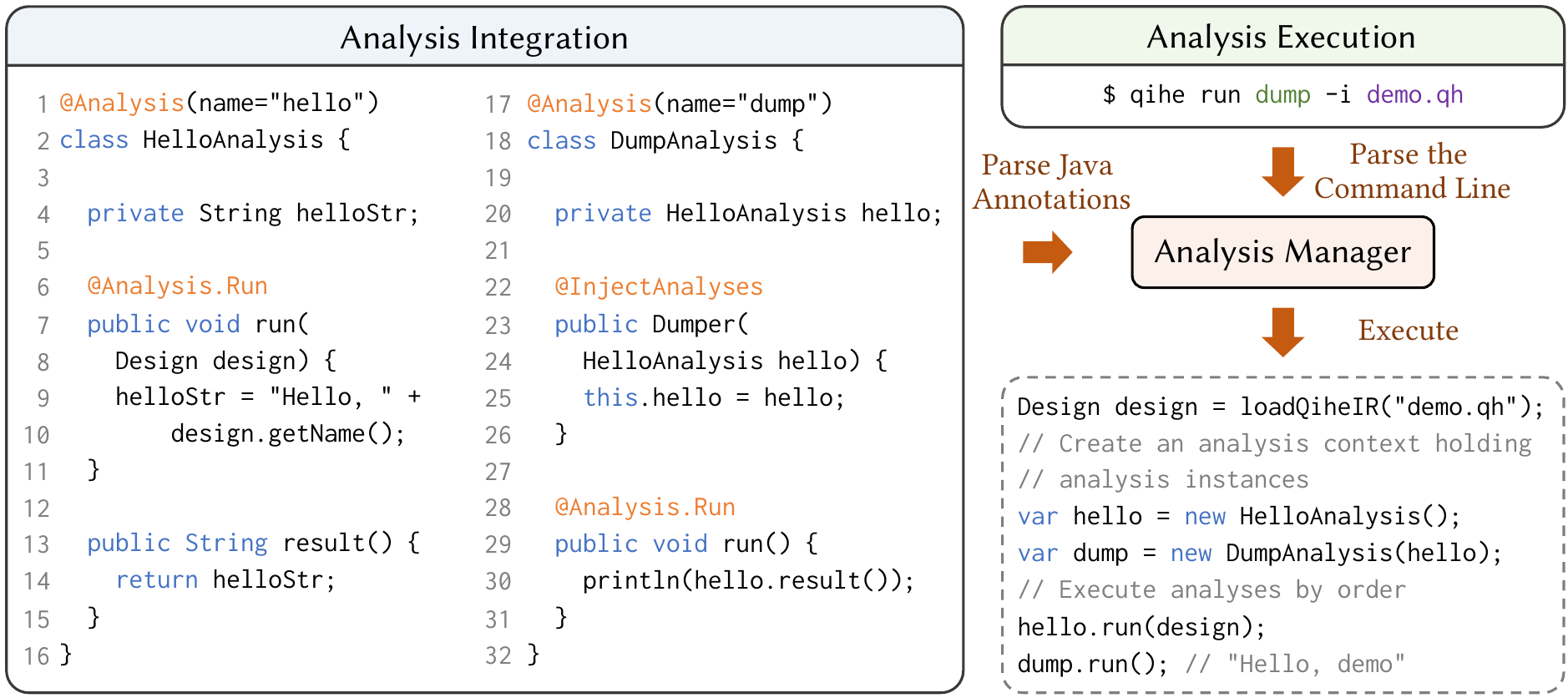}
\caption{An illustration of the process for integrating and executing analyses using the analysis manager. Qihe users only need to develop the analysis code (shown in "Analysis Integration") and specify which analyses to run (shown in "Analysis Execution").
The manager automatically executes the analyses in a manner similar to the code shown in the bottom-right of the figure.
For simplicity, the analysis context and analysis executor, two core components of the manager depicted in Figure~\ref{fig:arch}, are omitted here, and their logic is represented by the semantically equivalent code.}
\label{fig:manage}
\end{figure}

\subsection{Analysis Integration}

To simplify the integration of new analyses, Qihe offers annotation-based facilities to address the following common needs in analysis integration:
\begin{itemize}
\item Registering the new analysis for management.
\item Accessing the program abstraction to implement the analysis logic.
\item Exporting results for use by other analyses.
\item Retrieving results from existing analyses, if needed.
\end{itemize}
We will refer to Figure~\ref{fig:manage} to illustrate how each of these needs is fulfilled.
The code shown in Figure~\ref{fig:manage} is written in Java, as Qihe is implemented in it.

To address the first need (i.e., analysis registration), each analysis in Qihe must be a Java class annotated with \texttt{@Analysis(name="str")} (see lines~1 and~17 in Figure~\ref{fig:manage}).
This annotation enables the framework to automatically discover and register analyses.
The \texttt{name} attribute in the annotation serves as a brief identifier for the analysis, allowing users to specify which analysis to run via the command line.
For instance, Figure~\ref{fig:manage} defines two analyses named \aname{hello} and \aname{dump}, and uses the command
``\texttt{qihe run dump -i demo.qh}''
to execute the analysis \aname{dump} on the input Qihe IR \texttt{demo.qh}, as specified by the \texttt{-i} flag.
This IR file represents the Verilog design compiled from the Verilog source file \texttt{demo.v} (shown in Figure~\ref{fig:ir}) using the command ``\texttt{qihe compile demo.v -o demo.qh}''.
It is worth noting that, in real-world projects, Verilog designs are typically described across multiple files.
The ``\texttt{qihe compile}'' command can compile these files together into a single Qihe IR, which can then be analyzed using ``\texttt{qihe run}''.

To access the program abstraction (i.e., Qihe IR) and implement analysis logic, developers must define a \texttt{Run} method for each analysis (see lines~7 and~29).
The first parameter of this method is a \texttt{Design} object provided by Qihe, which represents the Qihe IR specified by the \texttt{-i} flag on the command line (e.g., \texttt{demo.qh} in this example).
If an analysis does not require direct access to the IR and only needs results from other analyses---for instance, a data-flow analysis that only relies on the control-flow graph provided by another analysis---this parameter can be omitted (as in the \texttt{dump} analysis).
The body of the \texttt{Run} method contains the analysis logic. 
The method must be annotated with \texttt{@Analysis.Run} so that Qihe can automatically identify and invoke it during analysis execution.

To export the results of an analysis, developers simply provide public methods that return the analysis results.
For example, in line~13 of Figure~\ref{fig:manage}, the \aname{hello} analysis defines a public method, \texttt{result}, which returns the ``Hello'' string computed in its \texttt{Run} method.

To access the results of existing analyses, a new analysis must declare its dependencies using a constructor that takes other analysis instances as parameters (see lines~22--26).
This constructor should be annotated with \texttt{@InjectAnalyses}, allowing Qihe to identify and invoke it to inject the dependencies.
If there are no dependencies, the constructor can be omitted (as in the \texttt{hello} analysis).
The new analysis can store these dependent analysis objects as fields and use their results in its \texttt{Run} method. 
For example, the \aname{dump} analysis retrieves the \texttt{``Hello''} string from the \aname{hello} analysis stored in a field and prints it to standard output.

\subsection{Analysis Execution}

A usual approach to executing analyses involves manually creating instances of analysis classes and invoking their \texttt{Run} methods in the correct dependency order.
For instance, to run the analysis \aname{dump}, as shown in the bottom-right of Figure~\ref{fig:manage}, one might load the Qihe IR, create instances of \aname{hello} and then \aname{dump}, and sequentially invoke their \texttt{Run} methods, passing the \texttt{Design} object if necessary.
\looseness=-1

However, as the number of analyses grows, manually managing this process becomes cumbersome and difficult to maintain. To address this issue, Qihe automates the entire process.
Users simply need to execute a command like:
\begin{verbatim}
                        qihe run <analysis> -i <ir>
\end{verbatim}
The framework then performs the following steps, achieving the same effect as generating and executing the code shown in the bottom-right of Figure~\ref{fig:manage} using Java's reflection mechanism:
\begin{enumerate}
\item
Loads the Qihe IR under analysis (i.e., \texttt{<ir>}) into a \texttt{Design} object.
\item
Scans the Qihe codebase for classes annotated with \texttt{@Analysis} to discover all available analyses and parses their annotations to record metadata.
\item 
Resolves the dependencies of these analyses, instantiates them, and injects their dependencies in order via constructors annotated with \texttt{@InjectAnalyses}.
If circular dependencies are detected, an error is reported.
\item 
Executes the requested \texttt{<analysis>} and its dependencies by invoking their \texttt{Run} methods in the correct dependency order, passing the \texttt{Design} object as needed.
Each \texttt{Run} method is invoked only once to avoid redundant computation.
\end{enumerate}

With Qihe managing these tasks, developers can focus on utilizing existing analyses to implement their analysis logic, leaving the complexity of runtime management to the framework.

\section{Evaluation}\label{sec:usefulness}

The purpose of Qihe is to support and facilitate the development of a wide range of static analysis clients in hardware.
To achieve this, Qihe leverages a snowballing process, where fundamental analyses synergistically combine to build increasingly sophisticated analyses, ultimately enabling diverse application-specific analyses.
To our knowledge, no prior work has explored these fundamental analyses or investigated their synergistic, snowballing-like composition.

This section begins with a case study (Section~\ref{sec:showcase}) that illustrates how the snowballing process enables the detection of critical missing-reset bugs by revealing the internal statistics integral to this process.
Subsequently, we evaluate Qihe's utility in supporting useful, application-specific analyses across diverse domains: hardware bug detection (Section~\ref{sec:bug}), hardware security analysis (Section~\ref{sec:security}), and hardware program understanding (Section~\ref{sec:understanding}).

\subsection{Snowballing Process in Qihe: A Case Study} \label{sec:showcase}

The growth of analysis capabilities in Qihe resembles the snowballing process, where existing analyses collaboratively contribute to increasingly sophisticated analyses.
This process ultimately allows us to detect bugs that were previously unattainable with existing static analyses~\cite{slang,svlint,verible,verilator} for Verilog bug detection, as summarized in Table~\ref{tab:bugs}.
Section~\ref{sec:supporting-case-study} has discussed an example of this snowballing process, where key analyses interdepend and collaborate to support the \aname{missing-reset} analysis.
In this section, we take a deeper look at this process by examining the experimental results associated with \aname{missing-reset}.
Specifically, we apply the \aname{missing-reset} analysis and its dependencies to real-world programs, emphasizing three key aspects from the experimental results: (1) each analysis contributes to downstream effectiveness (i.e., analysis soundness and precision), (2) our analysis suite significantly reduces development costs, and (3) the efficiency of fundamental analyses dominates overall system efficiency.

\begin{table}[tp]
\centering
\fontsize{7pt}{8pt}\selectfont
\caption{Results of \aname{missing-reset} and its representative dependencies, ordered topologically by their dependencies, across 13 real-world Verilog programs.
Columns represent programs, rows represent analyses (except the \texttt{Meta} row lists lines of code and IR of these programs), and cells show summarized statistics.
For \aname{cfg} and \aname{fi-def-chain} that provide graphical program abstraction, complete ground truth is infeasible to collect and thus omitted.
For \aname{clocks}, \aname{regs}, and \aname{resets}, results are measured by precision and recall against the ground truth (\#Truth), where precision measures the proportion of identified results that
are ground truth and recall indicates the proportion of ground truth results that are identified.
For \aname{missing-reset}, since no ground truth exists, we present how many reported bugs are confirmed by developers in the format ``\#Reported (\#Confirmed)''. If no bugs are confirmed, we do not count them as true bugs and omit ``(\#Confirmed)'', as our goal is to highlight how many reported bugs are actually of concern to developers.\looseness=-1}
\label{tab:analyses-results}
\resizebox{\textwidth}{!}{
\begin{tabular}{ccccccccccccccc}
\toprule
 &  & XS & ziu & biv & FPC & ha3 & riv & ax1 & sev & pi2 & dav & adf & st1 & sha\\
\midrule
\multirow{2}{*}[-.3em]{\texttt{Meta}} & LoC & 1,821,858 & 22,290 & 13,601 & 7,751 & 7,662 & 6,832 & 6,804 & 3,461 & 3,049 & 2,753 & 2,724 & 2,553 & 2,171\\
\cmidrule{2-15}
 & LoIR & 8,261,716 & 26,554 & 25,435 & 41,065 & 16,053 & 12,309 & 28,151 & 5,825 & 5,747 & 3,592 & 7,128 & 9,429 & 2,281 \\
\midrule
\multirow{2}{*}[-.3em]{\colorbox{clctrl}{\texttt{cfg}}} & \#Node & 5,020,594 & 15,308 & 14,762 & 18,641 & 8,130 & 7,035 & 17,159 & 3,362 & 3,170 & 2,013 & 4,215 & 5,844 & 1,846 \\
\cmidrule{2-15}
 & \#Edge & 4,355,660 & 14,646 & 13,773 & 24,735 & 8,583 & 6,635 & 15,812 & 2,892 & 3,398 & 1,799 & 4,161 & 5,100 & 1,932\\
\midrule
\multirow{2}{*}[-.3em]{\colorbox{cldata}{\texttt{fi-def-chain}}} & \#Node & 1,549,502 & 6,369 & 5,784 & 11,280 & 4,272 & 3,024 & 7,753 & 1,211 & 1,932 & 1,024 & 1,476 & 2,124 & 1,656 \\
\cmidrule{2-15}
 & \#Edge & 3,176,863 & 12,009 & 10,387 & 15,147 & 7,486 & 4,653 & 13,575 & 1,694 & 9,414 & 1,628 & 2,106 & 3,479 & 87,631 \\
\midrule
\multirow{3}{*}[-.7em]{\colorbox{clundstd}{\texttt{clocks}}} & \#Truth & 2,651 & 29 & 36 & 13 & 10 & 14 & 39 & 13 & 3 & 9 & 23 & 16 & 1 \\
\cmidrule{2-15}
 & Precision & 100\% & 100\% & 100\% & 87\% & 100\% & 100\% & 100\% & 100\% & 100\% & 100\% & 100\% & 100\% & 100\% \\
\cmidrule{2-15}
 & Recall & 100\% & 100\% & 100\% & 100\% & 100\% & 100\% & 100\% & 100\% & 100\% & 100\% & 100\% & 100\% & 100\% \\
\midrule
\multirow{3}{*}[-.7em]{\colorbox{clundstd}{\texttt{regs}}} & \#Truth & 63,678 & 367 & 197 & 152 & 97 & 153 & 604 & 50 & 156 & 52 & 66 & 263 & 32 \\
\cmidrule{2-15}
 & Precision & 100\% & 99\% & 86\% & 100\% & 100\% & 100\% & 100\% & 100\% & 99\% & 100\% & 100\% & 100\% & 100\% \\
\cmidrule{2-15}
 & Recall & 100\% & 100\% & 97\% & 100\% & 100\% & 99\% & 99\% & 100\% & 97\% & 100\% & 94\% & 100\% & 100\% \\
\midrule
\multirow{3}{*}[-.7em]{\colorbox{clundstd}{\texttt{resets}}} & \#Truth & 2,218 & 37 & 36 & 14 & 8 & 14 & 38 & 7 & 4 & 10 & 19 & 6 & 1 \\
\cmidrule{2-15}
 & Precision & 98\% & 97\% & 100\% & 92\% & 80\% & 100\% & 100\% & 100\% & 100\% & 50\% & 100\% & 100\% & 100\% \\
\cmidrule{2-15}
 & Recall & 88\% & 81\% & 100\% & 79\% & 100\% & 100\% & 100\% & 86\% & 75\% & 100\% & 100\% & 83\% & 100\% \\
\midrule
\colorbox{clbug}{\texttt{missing-reset}} & \#Reported & 50 & 9 & 0 & 0 & 0 & 0 & 20 (\textbf{2}) & 10 & 5 & 3 (\textbf{2}) & 1 (\textbf{1}) & 12 & 0 \\
\bottomrule
\end{tabular}

}
\end{table}

\paragraph{Program Set}
To provide practical statistics associated with the \aname{missing-reset} analysis, we select thirteen real-world hardware programs spanning diverse domains, such as CPU design, encryption, AXI protocols, and computer vision.
These programs are drawn from popular open-source GitHub projects~\cite{xiangshan,zipcpu,biriscv,fpga-foc,riscv,serv,picorv32,darkriscv,vtr,hazard3}, which average approximately 1.9K stars, as well as from hardware bug benchmarks~\cite{asplos22bugbed-artifact,hackdac18}, offering additional bug-related statistics.
The top of Table~\ref{tab:analyses-results} lists these programs by shorthand names for brevity, as well as their metadata, including lines of code (LoC) and lines of IR (LoIR) after compilation.
All these programs are available in our project repository.
To facilitate future research that develops new analysis clients atop our infrastructure (e.g., Verilog front end and IR) and evaluates them on these popular open-source projects, our repository preprocesses these projects into forms suitable for immediate analysis, such as precompiled IR or single Verilog files.
This preprocessing saves developers from repeating the substantial compilation effort, as Verilog projects lack a standard build system and often require careful understanding of each project's structure and build process to generate the IR.
\looseness=-1

\subsubsection{Representative Analyses and Results}

To showcase how analyses in Figure~\ref{fig:missing-reset-dep} enhance downstream effectiveness in the snowballing process, Table~\ref{tab:analyses-results} presents representative analyses and their results, organized in topological order.
The absence of certain analyses from this table, such as \aname{hierarchy} (for module hierarchy resolution in Verilog) and \aname{reaching-guards} (for synchronization information extraction), does not diminish their importance.
They are excluded because their results are either unsuitable for concise presentation in the table or require substantial human effort to collect their ground truth.
Nevertheless, their effectiveness is still reflected in the performance metrics of analyses listed in Table~\ref{tab:analyses-results}, as these analyses depend on them.

\aname{cfg} (control-flow graphs) and \aname{fi-def-chain} (flow-insensitive def-use chains) appear first in the table, highlighting their foundational role in enabling subsequent key analyses.
For example, the \aname{clocks} analysis, conducted later, utilizes edges from the \aname{fi-def-chain} to propagate clock signal information.
Initially, we explored a flow-sensitive implementation (\aname{fs-def-chain}) due to concerns that false positive edges in the \aname{fi-def-chain} might generate infeasible propagation paths, thereby reducing the precision of clock inference in \aname{clocks} analysis.
However, we discovered that the flow-insensitive version provides sufficient precision and significant speed benefits in practice, making it our preferred approach.
This case also implies that Qihe actually offers a range of fundamental analyses, allowing developers to explore alternatives to make preferred trade-offs in practice.
\looseness=-1

The next three analyses in the table (\aname{clocks}, \aname{resets}, and \aname{regs}) infer the physical usages of variables.
Their losses in precision and recall (the definitions of which are explained in the caption of Table~\ref{tab:analyses-results}) can significantly compromise the effectiveness of their dependent analyses and the practical utility in real-world scenarios, as discussed in Section~\ref{sec:supporting-case-study}.
Therefore, we dedicated careful effort for good precision and recall, and the results align with our expectations, as shown in Table~\ref{tab:analyses-results}.
Since these analyses are intended for hardware understanding, Section~\ref{sec:understanding} provides a detailed explanation of how we obtain the ground truth results and how these analyses achieve high precision and recall.

Finally, \aname{missing-reset} effectively identifies missing-reset bugs in real-world programs that could lead to unintended hardware behavior, leveraging the effectiveness of the preceding analyses.
As shown in Table~\ref{tab:analyses-results}, bold values indicate five bugs confirmed by developers.
It is worth noting that the number of reported missing-reset bugs remains manageable.
For example, even in the million-line XS program, only 50 bug reports are generated, which is acceptable for manual inspection.

Controlling the number of false bug reports is critical in hardware bug detection, as excessive reports can overwhelm developers and obscure true issues.
We address this in two ways:
(1)~By inspecting false positives in real-world programs, we introduce refinements tailored to each client when necessary, reducing spurious bug reports without significantly sacrificing recall.
For example, in the \aname{missing-reset} analysis, we skip excessively large cycles when detecting cycles in the dependency graph.
Such cycles are typically caused by false positive edges, which are inevitably introduced when building a sound dependency graph, and rarely correspond to genuine bugs.
(2)~We employ a general strategy for hardware analyses:
Verilog programs often use meta-programming to describe repetitive circuit patterns.
After elaboration, which expands meta-programming constructs, the same issue---especially a false positive---may be reported multiple times.
To address this, we collapse near-duplicate reports and emit a single representative.
While this approach may merge multiple instances of a true bug, fixing the representative root cause typically eliminates all related occurrences.
As a result of these efforts, the numbers of bugs (i.e., the bugs in Table~\ref{tab:bugs}) reported by our detection clients on real-world programs are \emph{all} manageable, similarly to the situation with \aname{missing-reset}, making it easier for users to validate the true bugs among them.

\subsubsection{Development Effort}
\begin{wrapfigure}[17]{r}{0.42\textwidth}
\centering
\vspace{-2.8ex}
\includegraphics[width=\linewidth,trim={10 10 10 10},clip]{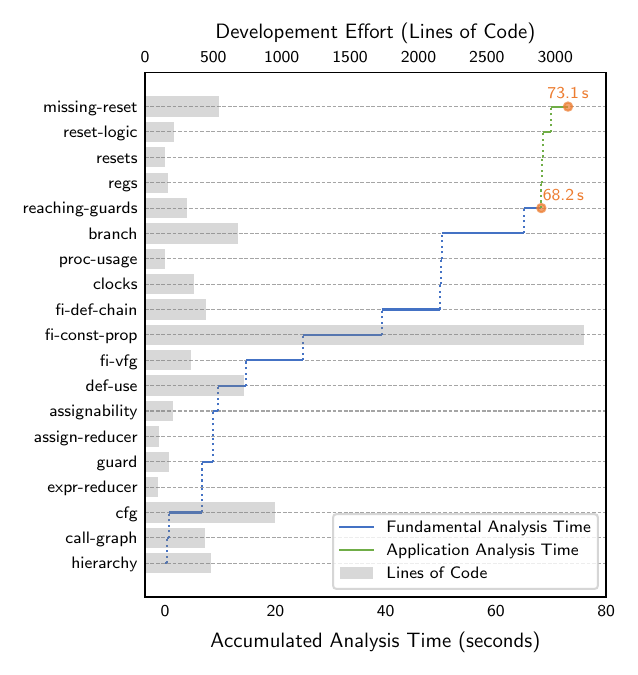}
\vspace{-5ex}
\caption{Missing-reset analysis on a large-scale RISC-V SoC (XS from Table~\ref{tab:analyses-results}).}
\label{fig:analysis-cost-chart}
\end{wrapfigure}

In the snowballing process, developing a new client becomes significantly easier by leveraging Qihe's fundamental analyses. 
The left axis of Figure~\ref{fig:analysis-cost-chart} shows all analyses contributing to \aname{missing-reset}, arranged from bottom to top in topological order based on their dependencies.
The gray bars represent the lines of code (LoC) for implementing each analysis.
As shown, implementing \aname{missing-reset} as a client within Qihe requires significantly less development effort (540~LoC) compared to implementing it from scratch without utilizing Qihe's fundamental analyses (9,700~LoC, calculated by summing the lines of code of all the required analyses for \aname{missing-reset}).
On average, each bug detection client built on our suite requires about 350~LoC, versus 5,900~LoC from scratch.
This substantial reduction in development effort underscores Qihe's ability to facilitate the development of new application-specific analyses.

\subsubsection{Analysis Time}\label{par:speedup}
Fundamental analyses, which address essential analysis needs with a focus on hardware characteristics, also shoulder the majority of the analysis workload required to support downstream analyses.
Figure~\ref{fig:analysis-cost-chart} shows the accumulated runtime cost of each analysis when running \aname{missing-reset} on XS, a large-scale RISC-V SoC from Table~\ref{tab:analyses-results} with 1.8M lines of code, measured on a machine with an Intel i9-13905H CPU and using 16\,GB of memory.
In this case, fundamental analyses (from \aname{hierarchy} to \aname{reaching-guards}) contribute 93\% of the total time (68.2\,s of 73.1\,s).
Similarly, across the other twelve programs from Table~\ref{tab:analyses-results}, they average 84\% of the total time.

This highlights the dominant role of fundamental analyses in Qihe, further indicating that optimizing fundamental analyses offers a significant opportunity to reduce the overall analysis time.
We accelerate these analyses by leveraging Verilog-specific properties---such as the acyclicity of combinational logic, the branchless nature of continuous assignments, and the synchronicity of guard statements---alongside conventional software optimization techniques.
Before optimizations, some analyses on XS took up to an hour using 128\,GB of memory.
After optimization, the \emph{entire} analysis suite (i.e., all 20 application-specific analyses, along with the 22 fundamental analyses in Figure~\ref{fig:analyses}) on XS completes in about 5 minutes using 40\,GB of memory.
For the other twelve programs in Table~\ref{tab:analyses-results}, totaling 82K lines, the entire suite completes in just 6 seconds.
Note that running multiple analysis clients together in one process enables the sharing of analysis results and reduces recomputation, but increases peak memory usage by about 1\,GB per additional analysis on this million-line XS program.
Therefore, running \aname{missing-reset} (involving 19 analyses) on XS takes about 1 minute and 16\,GB, whereas executing the entire suite (involving 42 analyses) completes in about 5 minutes using 40\,GB.
\looseness=-1

\subsection{Hardware Bug Detection}\label{sec:bug}

\begin{table}[t]
    \centering
    \caption{
        Hardware bugs identified by Qihe in real-world projects.
        All listed bugs are undetectable by existing linter-style static analyses for Verilog bug detection~\cite{slang,svlint,verible,verilator}.
        Newly discovered bugs are shown in \textbf{bold}; they were previously unknown and have been confirmed by developers.
    }
    \label{tab:bugs}
    \footnotesize
    \resizebox{\textwidth}{!}{
        \begin{tabular}{cclcl}
    \toprule
    Category & ID & Description & Project & Freshness \\
    \midrule
    \multirow{5}{*}[-.7em]{Missing Reset}
    & B01 & Register \texttt{drop\_frame} lacks proper reset logic. & \multirow{2}{*}{Verilog-AXIS~\cite{verilog-axis}} & \multirow{2}{*}{Known from~\cite{asplos22bugbed}.}\\
    & B02 & Register \texttt{wr\_ptr\_cur} lacks proper reset logic. & & \\
    \cmidrule{2-5}
    & B03 & Register \texttt{correct} lacks proper reset logic. & PULPissimo~\cite{PULPissimo} & Known from~\cite{tcad22rtlcon}.\\
    \cmidrule{2-5}
    & \textbf{B04} & Register \texttt{TIMER} lacks proper reset logic. & \multirow{2}{*}{DarkRISCV~\cite{darkriscv}} & \multirow{2}{*}{Newly discovered.}\\
    & \textbf{B05} & Register \texttt{TIMEUS} lacks proper reset logic. & & \\

    \cmidrule{1-5}
    \multirow{4}{*}[-.5em]{\makecell{Unreachable\\ States}}
    & B06 & Used state \texttt{dma\_ctrl\_reg == CTRL\_ABORT} is unreachable. & OpenPiton~\cite{OpenPiton} & Known from~\cite{hackdac21}. \\
    \cmidrule{2-5}
    & \textbf{B07} & Used state \texttt{count == 2} is unreachable. & \multirow{3}{*}{32-Verilog-Projects~\cite{verilogprojects}} & \multirow{3}{*}{Newly discovered.} \\
    & \textbf{B08} & Used state \texttt{count == 8} is unreachable. & & \\
    & \textbf{B09} & Used state \texttt{count == 32} is unreachable. & & \\

    \cmidrule{1-5}
    \multirow{2}{*}[-.3em]{Deadlock}
    & B10 & \texttt{o\_sclk} gets stuck from a deadlock. & SDSPI~\cite{sdspi} & \multirow{2}{*}[-.3em]{Known from~\cite{asplos22bugbed}.}\\
    \cmidrule{2-4}
    & B11 & \texttt{rd\_addr} gets stuck from a deadlock. & HDL Lib~\cite{hdllib} & \\

    \cmidrule{1-5}
    \multirow{3}{*}[-.7em]{Undriven Signals}
    & \textbf{B12} & Use of values from the undriven signal \texttt{clock\_div}. & Basic Verilog~\cite{basicverilog} & \multirow{3}{*}[-.7em]{Newly discovered.}\\
    \cmidrule{2-4}
    & \textbf{B13} & Use of values from the undriven signal \texttt{r\_cv}. & ZipCPU~\cite{zipcpu} & \\
    \cmidrule{2-4}
    & \textbf{B14} & Use of values from the undriven signal \texttt{v}. & 32-Verilog-Projects~\cite{verilogprojects} &\\

    \cmidrule{1-5}
    Unloaded Signals
    & B15 & Module \texttt{aes\_1cc} has an unloaded signal \texttt{rst} & PULPissimo~\cite{PULPissimo} & Known from~\cite{tcad22rtlcon}.\\

    \cmidrule{1-5}
    Submodule Misuse
    & \textbf{B16} & Misuse of submodule \texttt{invert} due to wrong port order. & 32-Verilog-Projects~\cite{verilogprojects} & Newly discovered.\\

    \cmidrule{1-5}
    Mis-Truncation
    & B17 & Mis-truncation of \texttt{rx\_mmio\_channel} causes data loss. & Sha512~\cite{sha512} & \multirow{2}{*}[-.3em]{Known from~\cite{asplos22bugbed}.}\\

    \cmidrule{1-4}
    Invalid Use
    & B18 & Use of invalid value from signal \texttt{axi\_adrv9001\_rx\_channel}. & HDL Lib~\cite{hdllib} & \\
    \bottomrule
\end{tabular}

    }
\end{table}

Detecting hardware bugs remains a formidable challenge, even in industrial settings.
According to Siemens' comprehensive global industrial study in 2024~\cite{wilson24}, merely 14\% of IC/ASIC projects achieved first silicon success, and only 13\% of FPGA projects reported zero bug escapes into production.
Unlike software, where bugs can be patched remotely over the internet at a low cost, late-stage hardware bugs in production can result in exponentially increasing remediation expenses~\cite{dvcon16cost, dac04costs}.
\looseness=-1

Static analysis offers a lightweight approach for early-stage bug detection at design time, making it especially cost-effective for hardware.
To showcase Qihe's capability as a general-purpose framework for supporting hardware bug detection clients, as described in Section~\ref{sec:analyses}, we developed a set of Verilog analyses, the insights behind which were derived from studying known bugs in the most recent hardware bug survey~\cite{asplos22bugbed,asplos22bugbed-artifact}, the renowned annual hardware bug detection hackathon~\cite{hackdac,hackdac18,hackdac21}, and through direct communication with industrial hardware practitioners~\cite{hisilicon,t-head}.

As shown in Table~\ref{tab:bugs}, the preliminary experimental results are very promising.
Our analyses successfully identified 18 bugs across 8 categories---such as missing resets, unreachable states, and deadlocks---in a diverse set of real-world hardware projects, such as system-on-chips (SoCs), communication protocols, and encryption modules.
As confirmed by our attempts, none of these bugs could be detected by existing Verilog linters~\cite{slang,svlint,verible,verilator}, the dominant static analysis tools for hardware bug detection.
Their syntax-based approach and lack of fundamental semantic analysis prevent them from identifying bugs that require deeper hardware semantic reasoning.
Notably, 9 of the 18 bugs were newly uncovered by our analyses from popular open-source projects~\cite{darkriscv,basicverilog,zipcpu,verilogprojects} (averaging over 1.5K GitHub stars).
All of these new bugs have since been confirmed by the respective developers.
These results not only highlight the effectiveness of our analyses but also underscore the promise of sophisticated static analysis for practical hardware bug detection.
\looseness=-1

To further illustrate how our analyses identify additional categories of real-world hardware bugs beyond the missing-reset bugs already discussed in Section~\ref{sec:motivating-example}, we present two additional case studies---an unreachable-state bug (B09) and a hardware deadlock bug (B10)---as supplementary representative examples of sophisticated Verilog static analyses for practical hardware bug detection.

\subsubsection{An Unreachable-State Bug Case Study}

\begin{figure}[tp]
    \begin{subfigure}[b]{.43\textwidth}
        \centering
        \includegraphics[trim=7.85cm 4.8cm 0cm 0,clip,page=1]{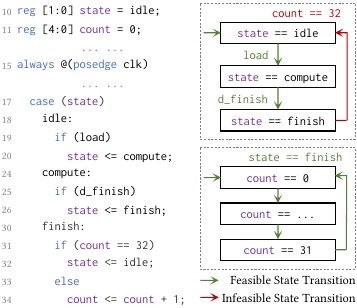}
        \caption{
            A real-world unreachable-state bug and its finite-state machine (FSM) diagram.
        }\label{fig:bug-unreachable}
    \end{subfigure}
    \hfill
    \begin{subfigure}[b]{.54\textwidth}
        \centering
        \includegraphics[trim=0 4.8cm 6.4cm  0,clip,page=1]{figures/bugs.pdf}
        \caption{
            A real-world hardware deadlock bug and its process dependency graph.
        }\label{fig:bug-deadlock}
    \end{subfigure}
    \vspace{-2ex}
    \caption{
        Real-world bug case studies: B09 (\subref{fig:bug-unreachable}) and B10 (\subref{fig:bug-deadlock}) from Table~\ref{tab:bugs}.
        Line numbers correspond to those in the original source projects.
    }
    \vspace{-2ex}
\end{figure}

Finite-state machine (FSM) design is a common task for Verilog engineers~\cite{fsm}.
An unreachable state is one that the FSM cannot reach from any initial state.
Such states often indicate functional issues, such as unintended infeasible state transitions~\cite{unreachable}.

Figure~\ref{fig:bug-unreachable} presents a real-world unreachable-state bug from a CRC-32 serial counter in an open-source Verilog project~\cite{verilogprojects}, where \texttt{idle}, \texttt{compute}, and \texttt{finish} are constant enumeration values for \texttt{state}.
The diagram on the right visualizes two FSMs defined by the Verilog code on the left: nodes represent states (corresponding to the values of \texttt{state} and \texttt{count}, respectively), edges denote transitions, and expressions on edges indicate transition conditions.
Since all transitions in the \texttt{count} FSM (the lower one) share the same condition, we show it only once to avoid redundancy.
The bug arises as follows.
Lines~30--34 of Figure~\ref{fig:bug-unreachable} describe the interaction between the \texttt{state} FSM (the upper one) and the \texttt{count} FSM:
when \texttt{state} is \texttt{finish}, \texttt{count} starts incrementing, and when \texttt{count == 32}, \texttt{state} should return to \texttt{idle}.
However, the developer failed to notice that \texttt{count} is declared as a 5-bit vector (line~11) that overflows after reaching 31 and wraps back to 0, an easy-to-overlook case in Verilog programming.
Consequently, the state where \texttt{count == 32} is unreachable in the \texttt{count} FSM, breaking the intended interaction between the two FSMs and causing the \texttt{state} FSM to exhibit an infeasible transition from \texttt{state == finish} to \texttt{state == idle} (highlighted in red).
This flaw is severe, as it ultimately traps the system in the \texttt{finish} state and prevents it from ever resuming normal operation.

\begin{wrapfigure}[6]{r}{2cm}
\vspace{-2.5ex}
\hasse
\vspace{-2ex}
\end{wrapfigure}
To detect such unreachable states, our analysis performs \emph{Verilog-specific bit-level} whole-program constant propagation to determine all possible values of every variable with bit-level precision.
As discussed in Section~\ref{sec:supporting-case-study}, this precision is critical in Verilog, where programs frequently manipulate individual bits, and loss of precision can cause the analysis to miss bugs that would otherwise be detected, such as the one in this case study.
For each variable, our constant propagation overapproximates its concrete bit-vector value using a specially designed abstract bit vector, where each bit is drawn from the lattice shown on the right:
$\mathtt{U}$ represents an undefined bit,
$\mathtt{0}$/$\mathtt{1}$/$\mathtt{X}$/$\mathtt{Z}$ represent the fixed Verilog bit values 0/1/X/Z,
$\mathtt{B}$ (short for \emph{both}) represents a bit that can be either $\mathtt{0}$ or $\mathtt{1}$, and
$\mathtt{T}$ represents the top lattice element encompassing all possible bit values.

In this case study, we focus primarily on the $\mathtt{0}$/$\mathtt{1}$/$\mathtt{B}$ portion;
for example, for a three-bit signal $x$, $x = \mathtt{B01}$ implies that $x$ can take only two possible values: $\mathtt{001}$ or $\mathtt{101}$.
In Figure~\ref{fig:bug-unreachable}, the transition condition \texttt{count == 32} (line~31) is interpreted by our analysis as comparing $\mathtt{0BBBBB}$---with \texttt{count} zero-extended to 6 bits (an implicit type conversion per the Verilog specification~\cite{verilog-std-2005})---against $\mathtt{100000}$, representing the fixed value 32.
This equality is impossible because the most significant bit never matches ($0 \ne 1$);
therefore, the state where \texttt{count == 32}, used in lines 31--32 of Figure~\ref{fig:bug-unreachable}, is unreachable.
In contrast, directly using a conventional non-bit-level constant propagation would treat the left-hand side as a non-constant value (denoted $\mathtt{NAC}$), so evaluating $\mathtt{NAC} == 32$ would yield a ``possible'' (reachable) rather than ``impossible'' (unreachable) result, causing this unreachable-state bug to be missed.
This example underscores that directly applying software-like analyses to hardware is often ineffective, and that Verilog-specific customizations, such as maintaining bit-level precision, are essential.
Additional designs and algorithms for Verilog-specific static analysis are available in our open-source framework and documentation.

\subsubsection{A Hardware Deadlock Bug Case Study}

According to the Verilog language specification~\cite{verilog-std-2005}, a running Verilog program consists of multiple concurrent processes, each typically corresponding to a runtime instance of an \texttt{always} block.
A hardware deadlock occurs when circular dependencies among concurrent Verilog processes cause the program to stall infinitely~\cite{asplos22bugbed}, manifesting physically as signals that always fail to update as expected.

Figure~\ref{fig:bug-deadlock} presents a real-world deadlock bug from the \texttt{llsdspi} module of the SDSPI project~\cite{sdspi}, which implements an SD card controller.
This controller is expected to generate a driver clock signal named \texttt{o\_sclk}.
However, \texttt{o\_sclk} fails to update due to a deadlock caused by circular dependencies among three processes: \texttt{Process-147}, \texttt{Process-175}, and \texttt{Process-220}, where \texttt{Process-}$i$ denotes the runtime Verilog process corresponding to the \texttt{always} block starting at line~$i$.
Specifically, the update of \texttt{o\_sclk} (line~294) in \texttt{Process-220} relies on the update of \texttt{r\_z\_counter} (line~192) in \texttt{Process-175}, which in turn relies on the update of \texttt{startup\_hold} (line~152) in \texttt{Process-147}, which again relies on the update of \texttt{o\_sclk} (line~294) in \texttt{Process-220} through line~148.
As a result, all three processes remain infinitely stalled, preventing the SD card controller from generating the active driver clock \texttt{o\_sclk}.
Without this clock, the SD card becomes completely non-functional, unable to perform any data access operations.

To detect such hardware deadlock bugs, our analysis first constructs a \emph{process dependency graph} that captures whether the value updates in one Verilog process depend on those in another, as exemplified in Figure~\ref{fig:bug-deadlock}.
In this graph, an edge from \texttt{Process-}$i$ to \texttt{Process-}$j$ ($i \ne j$) labeled with~$x$ indicates that \texttt{Process-}$i$ is dependent on \texttt{Process-}$j$ through a shared signal~$x$; that is, $x$ is used in a branch condition within \texttt{Process-}$i$ and defined by an assignment in \texttt{Process-}$j$.
To identify circular dependencies among concurrent Verilog processes, our analysis detects all cycles in this graph.
For the example in Figure~\ref{fig:bug-deadlock}, the analysis reports a circular process dependency among \texttt{Process-147}, \texttt{Process-175}, and \texttt{Process-220}, pinpointing the deadlock that stalls the system.
\looseness=-1

\subsection{Hardware Security Analysis}\label{sec:security}

Hardware security has become increasingly crucial due to the widespread use of Integrated Circuits (ICs) in systems that are particularly vulnerable to security threats like sensitive information leaks~\cite{date21informationflow} and malicious hardware Trojans~\cite{informationflow_trojan}.
To further demonstrate Qihe's utility as a general framework to facilitate the development of various application-specific analyses, we developed two security analyses, \aname{taint} and \aname{x-prop}, as clients on top of Qihe.

\paragraph{Taint Analysis}

Qihe's \aname{taint} analysis is designed to address \textit{information flow security} concerns.
In this context, users specify \textit{taint sources} and \textit{sinks} to require no information flows from sources to sinks, as such a flow indicates potential vulnerabilities like unauthorized access or confidential information leaks.
Taint sources in Verilog could be given as registers holding confidential data, while taint sinks could be nets or module ports susceptible to exploitation by attackers.

Compared to information flow analysis in software, tracking information flow in hardware introduces additional complexities due to hardware-specific semantics, such as bit-level value propagation and bidirectional port connections.
Nevertheless, Qihe's infrastructure effectively handles and abstracts these hardware features, enabling the successful development of a static information flow analysis based on classical explicit flow modeling.

To evaluate Qihe's taint analysis (i.e., \aname{taint}), we utilized Trust-Hub~\cite{iccd13trusthub,jhss2017trusthub}, the de-facto standard benchmark widely used for assessing the effectiveness of hardware security verification techniques.
From its collection of Verilog designs, we collected all 25 related to information flow security and excluded 2 VHDL designs (incompatible with Qihe's Verilog-specific frontend) and 4 designs lacking explicitly specified taint sources/sinks.
This resulted in 19 Verilog designs for final evaluation.
These selected designs focus primarily on cryptographic hardware implementations, reflecting real-world applications where information flow security needs to be addressed.

Our analysis successfully identified 15 out of 19 information leak instances, where secret keys or the contents of internal state registers were leaked via observable output ports.
After manual inspection, the 4 cases not identified by \aname{taint} were implicit information flows caused by control dependencies.
Currently, implicit covert channels (e.g., control dependencies, delays, and power consumption) are not handled by \aname{taint}, but we believe it serves as a case for future work to feature more powerful analyses on top of Qihe.

\paragraph{X-Propagation Analysis}

\begin{figure}[tp]
\includegraphics[trim=0 330 100 0,clip,page=1,width=0.9\linewidth]{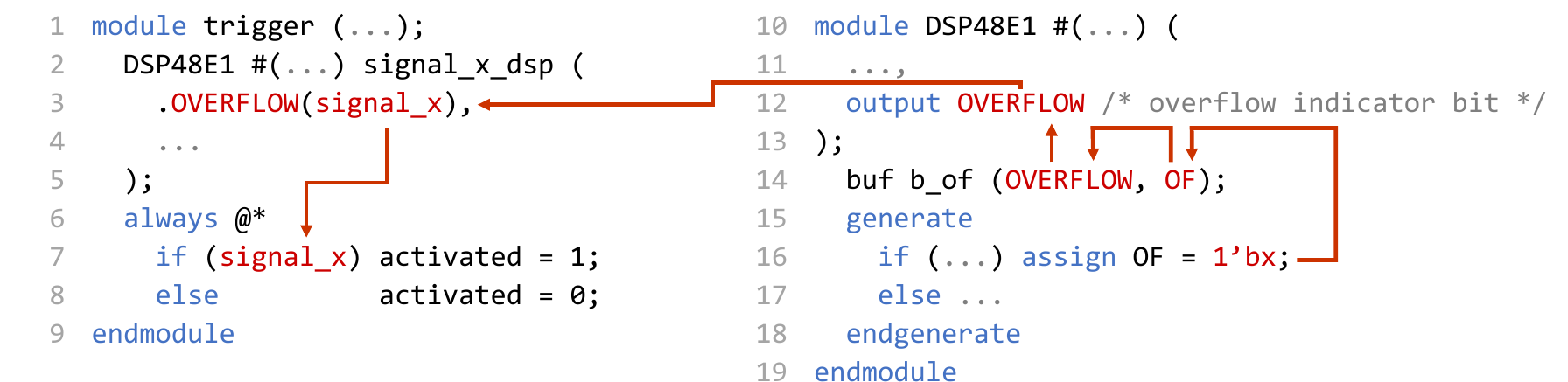}
\caption{A simplified excerpt of the hardware Trojan from~\cite{dac17mux} making use of X-Propagation. The value {\color{red}\texttt{1'bx}} is propagated through various logic gates and port connections, as shown in red direct edges, ultimately reaching an \texttt{if} statement, which serves as the exploit point.}\label{fig:xprop-example}
\end{figure}

Qihe's \aname{x-prop} analysis is designed to identify the propagation of \veriX-values (unknown states), which often signify critical design flaws.
These flaws can result in simulation failures or, more alarmingly, be exploited by hardware Trojans to conceal malicious behaviors.

Detecting \veriX-propagation issues is non-trivial as it requires careful handling of \veriX's unique semantics only seen in HDLs such as Verilog, and is further complicated when intertwined with hardware behaviors such as reset, memory, and bit addressing.
Rather than developing a standalone analysis that accounts for every detail to address such challenges, Qihe's \aname{x-prop} can focus on modeling \veriX-related semantics by leveraging existing fundamental analyses to divide and conquer the intricate hardware behaviors.
This approach results in a succinct implementation containing no more than 300 lines of code.
In our implementation, Qihe leverages \aname{resets} and \aname{missing-reset} to understand the hardware reset behavior; infers the possible value range of variables for out-of-bounds access detection via constant propagation from \aname{fi-const-prop}; captures bit-level semantics of Verilog by utilizing \aname{bit-refs}.

Figure~\ref{fig:xprop-example} illustrates a hardware Trojan from~\cite{dac17mux} identified by Qihe through its \texttt{x-prop} analysis.
The example has been simplified for ease of understanding.
We first explain how the Trojan is concealed by the \veriX value and then describe how \aname{x-prop} detects it.

To conceal the Trojan, the attacker exploits the discrepancy in the interpretation of the unknown value \veriX between simulation and synthesis, rendering the Trojan undetectable during simulation but triggerable in fabricated circuits. 
In lines 7--8 of Figure~\ref{fig:xprop-example}, the attacker employs specialized circuit logic (detailed in the following paragraph) to ensure that the condition \texttt{signal\_x} in the \texttt{if} statement always evaluates to \veriX.
During simulation, this \veriX value is treated as \texttt{false} as specified by the language standard~\cite{verilog-std-2005}, so \texttt{activated} remains \texttt{0}.
Since \texttt{activated} controls the execution of the Trojan payload---which could perform actions such as leaking secret keys (omitted for brevity in the example)---the payload is never triggered in simulation. 
However, after synthesis, the unknown \veriX value is resolved to either \texttt{0} or \texttt{1} in the physical circuit.
Thus, the condition \texttt{signal\_x} may evaluate to \texttt{1}, allowing \texttt{activated} to be set to \texttt{1} and enabling the Trojan payload.

To ensure that the \texttt{if} statement's condition evaluates to \veriX during simulation and is less likely to be detected, the attacker often hides the source of \veriX in a separate module. 
As highlighted by the red path in Figure~\ref{fig:xprop-example}, the attacker first configures a third-party DSP module (i.e., \texttt{DSP48E1}) to generate a \veriX value on the variable \texttt{OF} (line 16).
This \veriX value is then propagated through a buffer gate (line 14) to the output port \texttt{OVERFLOW} of the \texttt{DSP48E1} module (line 12).
From there, the \veriX value is further propagated into the \texttt{signal\_x} variable in the \texttt{trigger} module via the connection to the output port of the instantiated \texttt{DSP48E1} (lines 2--3).
Ultimately, \texttt{signal\_x} delivers the \veriX value to the condition of the \texttt{if} statement (line 8).
\aname{x-prop} successfully detects that the variable \texttt{OF} in \texttt{DSP48E1} may contain a \veriX value during simulation and traces the propagation path from \texttt{OF} to the \texttt{if} statement.
As a result, \aname{x-prop} flags this as a potential X-Propagation exploit and reports it as an error.
Through manual analysis, we confirmed the presence of a Trojan leveraging \veriX values for concealment purposes.

\subsection{Hardware Program Understanding}\label{sec:understanding}

Hardware design is a process involving multiple stages, such as functional design, synthesis, verification, etc., ultimately leading to the physical implementation.
When utilizing Verilog for functional design, it is crucial for hardware designers to grasp the details of physical implementation. For instance, they need to determine whether their clocks are integrated into complex logic to minimize latency and understand how registers are mapped to on-board resources to optimize resource utilization. Unfortunately, this vital information is often only disclosed after the lengthy synthesis process.

To help developers better understand their Verilog designs earlier, Qihe developed seven analyses for hardware program understanding that predict physical implementation details, with each analysis focusing on a specific understanding task.
For instance, the \aname{clocks} analysis predicts variables that function as part of the physical clock tree and flags improper usages like those used in combinational logic, while the \aname{regs} analysis infers variables mapped to physical registers.

Notably, Qihe completes these seven analyses in approximately one minute on the real-world program XS from Table~\ref{tab:analyses-results} comprising 1.8 million lines.
In contrast, synthesizing the same codebase using Yosys~\cite{yosys}, the most popular open-source synthesizer, exceeds our one-hour time budget.

Below, we examine the effectiveness of Qihe's analyses through two case studies: the \aname{clocks} analysis and the \aname{regs} analysis.
We reuse the 13 diverse real-world programs introduced in Section~\ref{sec:showcase}, and the results are already presented in Table~\ref{tab:analyses-results}.

In the table, the ground truth is taken from Yosys's results since our analyses aim to predict the post-synthesis physical implementation.
It is important to note that Yosys typically does not provide APIs to access results commonly required by static analysis, as it is not designed for this purpose.
For example, it does not provide APIs to dump the commonly needed clock tree in a circuit---a feature supported by our \aname{clocks} analysis.
This limitation thus complicates the process of building ground truth: to obtain the true clock tree, we had to modify Yosys to dump the clock tree leaves (i.e., clock signals directly connected to registers), extract the variables connected to these leaves to approximate the clock tree, and then manually verify their validity as clock tree signals.

As shown in Table~\ref{tab:analyses-results}, the effectiveness of these analyses is measured using precision and recall.
These metrics are critical for hardware program understanding, as insufficient performance in either metric would lead to excessive false positives or false negatives, compromising the tool's practical utility in real-world scenarios. 
Thus, we devoted careful effort to design algorithms that go beyond simplistic pattern matching to achieve high precision and recall, and the results, as shown in the table, align with our expectations.

For \aname{clocks}, it achieves excellent precision and recall across all evaluated programs due to our semantic-reasoning approach.  
The key insight is that hardware clocks form a tree structure whose boundaries with the rest of the circuit can largely be inferred from code semantics.
Once these boundaries are identified, the entire tree can be extracted, starting from any tree node.

The \aname{regs} analysis also achieves high precision and recall, leveraging the effectiveness of the \aname{clocks} analysis it depends on and its hardware semantic-level reasoning algorithm.
Its recall is slightly reduced in some programs because it misses registers updated via blocking assignments, rather than the standard non-blocking assignments.
While Yosys accommodates such cases, blocking assignments cannot express the register's ``update value in next clock cycle'' semantics~\cite{lambdav}, and it is widely recognized as bad practice to use blocking assignments for register updates.
Therefore, this minor loss of recall in such cases does not raise practical concerns for \aname{regs}.

To summarize, these quick and accurate analyses for program understanding tasks can offer developers effective insights in the early stage of hardware development, and hopefully, more such analyses can be produced to further enhance hardware programming efficiency in the future.

\section{Related Work}\label{sec:rela}
This work focuses on static analysis for hardware, with the most pertinent studies already discussed in Section~\ref{sec:intro}. 
As the first attempt to create a general-purpose static analysis framework for Verilog, we briefly elucidate below how our idea was shaped by previous research and highlight their limitations in achieving our objective.

Existing static analysis tools for hardware, such as linters~\cite{svlint,verible,slang}, have proven to be effective in production for simple bug detection and code style checking.
However, the rapid growth of hardware complexity and diversity has posed new challenges in reliability, security, and maintainability~\cite{meltdown,spectre,cacheattack,foreshadow}, which are far beyond the reach of simple syntax- or pattern-based static analyses, forcing developers to switch back to resource-intensive testing and verification~\cite{fmcad07ic3,tcad24modelchecking,usenixsecurity22fuzzing1,usenixsecurity22fuzzing2}.

This coincides with the history of static analysis for software quality: researchers and engineers first adopted testing and verification~\cite{modelchecking,hoare} with only simple static analyses for pattern-based checks~\cite{findingbugsiseasy,clint}, and later sophisticated static analyses for bug detection and security, even for million-line code~\cite{grasshopper,pinpoint,coverity,analysis4safety,svf,soot,wala,doop,taie}.
We believe the vast potential of sophisticated static analysis for addressing emerging challenges in hardware is yet to be explored.

This naturally raises the question: on what foundation can we build such sophisticated static analyses for hardware?
An intuitive solution would be to leverage existing tools that already incorporate frontends and IRs.
However, none of these tools are suited for developing sophisticated hardware static analyses~\cite{pyverilog,veripy,circt,llhd,svlint,verible,slang}.
This limitation stems from the fact that they are not specifically designed for static analysis and inevitably prioritize other design goals over analysis, let alone their lack of infrastructure, such as a set of necessary fundamental analyses to support diverse sophisticated analysis clients.
Therefore, we introduce Qihe---a general-purpose static analysis framework for Verilog that explicitly prioritizes the needs of hardware analysis in the design of all its components.

\section{Conclusion}

We introduce Qihe, the first general-purpose static analysis framework for Verilog, designed to support a wide variety of application-specific analyses.
This is accomplished through the collaboration of a comprehensive suite of fundamental analyses, an analysis-oriented front end and IR, and a facilitative analysis manager.
To showcase Qihe's potential, on top of its infrastructure, we further developed client analyses that successfully detect hardware bugs and vulnerabilities, and facilitate program understanding tasks, all in real-world projects. 
This work marks a preliminary step toward unlocking the full potential of sophisticated static analysis for hardware.
We hope that Qihe, with its over 100K lines of fully open-source code, could inspire researchers and engineers from both the software and hardware communities to work in collaboration in advancing static analysis for hardware in the future.

\begin{acks}
This project was undertaken out of curiosity and a passion for exploration.
We extend our gratitude to Xuanlin Li from HiSilicon (Huawei) for introducing us to the significance of static analysis in hardware. Additionally,
we thank Yufei Liang, Teng Zhang, Menglong Chen, and Zhiwei Zhang for their helpful suggestions that contributed to the improvement of this paper.
We also acknowledge Zhongyi Cai and Sitao Lin for their early use of our tool and their constructive feedback.

\end{acks}

\bibliographystyle{ACM-Reference-Format}
\bibliography{references}


\end{document}